\documentclass[onefignum,onetabnum]{siamonline190516}

\usepackage{amsfonts}
\usepackage{graphicx}
\usepackage{epstopdf}
\usepackage{algorithmic}
\ifpdf
  \DeclareGraphicsExtensions{.eps,.pdf,.png,.jpg}
\else
  \DeclareGraphicsExtensions{.eps}
\fi

\usepackage{bm}
\usepackage[shortlabels]{enumitem}

\usepackage[sort]{cite}

\graphicspath{{figures/}}

\usepackage{enumitem}
\setlist[enumerate]{leftmargin=.5in}
\setlist[itemize]{leftmargin=.5in}

\newsiamremark{remark}{Remark}
\newsiamremark{hypothesis}{Hypothesis}
\crefname{hypothesis}{Hypothesis}{Hypotheses}
\newsiamthm{claim}{Claim}

\headers{Swarmalators with higher harmonic coupling}{L. D. Smith}

\title{Swarmalators with higher harmonic coupling: Clustering and vacillating
}

\author{Lauren D Smith\thanks{Department of Mathematics, The University of Auckland, Auckland 1142, New Zealand 
  (\email{lauren.smith@auckland.ac.nz}).}
}

\usepackage{amsopn}

\ifpdf
\hypersetup{
  pdftitle={Swarmalators with higher harmonic coupling},
  pdfauthor={LD Smith}
}
\fi

\begin{document}

\maketitle

\begin{abstract}
We study the dynamics of a swarmalator model with higher harmonic phase coupling. We analyze stability, bifurcation and structural properties of several novel attracting states, including the formation of spatial clusters with distinct phases, and single spatial clusters with a small number of distinct phases. We use mean-field (centroid) dynamics to analytically determine inter-cluster distance. We also find states with two large clusters along with a small number of swarmalators that are trapped between the two clusters and vacillate (waver) between them. In the case of a single vacillator we use a mean-field reduction to reduce the dynamics to two-dimensions, which enables a detailed bifurcation analysis. We show excellent agreement between our reduced two-dimensional model and the dynamics and bifurcations of the full swarmalator model.

\end{abstract}

\begin{keywords}
  swarmalators, coupled oscillators, model reduction
\end{keywords}

\begin{AMS}
37N25, 34C20, 34D06
\end{AMS}

\section{Introduction}
While oscillatory dynamics \cite{BickEtAl2020, MontbrioEtAl2015, SchmidtAvitabile2020, MachowskiEtAl2011, NishikawaMotter2015, FilatrellaEtAl08, Kuramoto84, Strogatz00, PikovskyEtAl01, AcebronEtAl05, OsipovEtAl07, ArenasEtAl08, DorflerBullo14, RodriguesEtAl16, Laing2014, Laing2018, OmelchenkoLaing2022} and swarming dynamics \cite{Sumpter2010, Reynolds1987, VicsekEtAl1995, TopazBertozzi2004, BalleriniEtAl2008, FetecauEtAl2011, BialekEtAl2012, HemelrijkEtAl2012, BernoffTopaz2013, HerbertRead2016}  have been considered in detail separately, there have been comparatively few studies on the dynamics of so-called ``swarmalators'', which have bi-directionally coupled oscillatory and swarming dynamics \cite{Tanaka2007, IwasaEtAl2010, OKeeffeEtAl2017, OKeeffe2019, SarGhosh2022, SarEtAl2022}. Examples of swarmalators in nature include microswimmers such as sperm cells which aggregate and synchronize the beating of their flagella \cite{TungEtAl2017, ElfringLauga2009}, as well as myxobacteria \cite{IgoshinEtAl2001}. To date, only first-harmonic sinusoidal phase interactions have been considered. As a step toward considering general coupling functions, we extend the original swarmalator model \cite{OKeeffeEtAl2017} to include higher harmonic coupling in the phase dynamics. Since pairwise coupling is generally considered to be anti-symmetric (equal and opposite), and phase variables are $2\pi$-periodic, general phase coupling functions can be expressed as Fourier sine series. We consider truncation of such Fourier sine series to the most dominant modes. In particular, we focus on the dynamics that results from phase coupling functions such that the first and second harmonics are equally dominant, and then the dynamics that results from a single dominant higher harmonic.

We show that including second harmonic coupling yields many new attracting states, including the formation of spatially separated clusters, each having a single phase, and single cluster states with exactly two phases and a complex crystalline structure. We analyze the stability properties of these new states and determine the parameter regions in which they are stable. For the state with two spatially separated anti-phase clusters, which occurs when same-phase swarmalators are spatially attracted and opposite-phase swarmalators are repelled, we use a mean-field (centroid) reduction to obtain a simple analytical expression for the cluster separation distance. Our result is similar to that of Sar \textit{et al.} \cite{SarEtAl2022}, though the underlying dynamics are fundamentally different. For the state with a single spatial cluster and two phases, which occurs when same-phase swarmalators are spatially repelled and opposite-phase swarmalators are attracted, we analyze how well the two phases mix together. We show that as the strength of attraction and repulsion is increased, there is greater mixing between the swarmalators with distinct phases.




In addition to clustered states, we have discovered states with two large anti-phase spatial clusters along with a small number of swarmalators that are trapped between them. The trapped swarmalators vacillate (waver) between the clusters. We find that these states occur on one edge of the stability region for the two-cluster state. We derive reduced mean-field dynamics for the vacillators. In the case of a single vacillator, the dynamics is effectively two-dimensional, which allows a detailed bifurcation analysis. Our analysis shows a Hopf bifucation from stable stationary behavior to oscillatory dynamics, as well as a heteroclinic and homoclinic bifurcations that corresponds to the transition from oscillatory dynamics to being absorbed into one of the larger clusters. We demonstrate excellent agreement between our reduced model and the dynamics and bifurcations of the full model.

The paper is organized as follows: In Section~\ref{sec:model} the model with second harmonic coupling is introduced, then in Section~\ref{sec:static_single_cluster} the stability of a single spatial cluster with a single phase is analyzed. In Section~\ref{sec:2_phase_stability} the stability region for states with two distinct phases is determined. Section~\ref{sec:anti-phase_one_cluster} studies states with a single spatial cluster and two distinct phases, considering properties such as mixing of phases within the crystalline lattice. In Section~\ref{sec:anti-phase_clusters} we study states with two anti-phase spatial clusters, including their separation distance, and in Section~\ref{sec:vacillators} we study vacillator dynamics (swarmalators that waver between two large clusters). We extend our clustering results for higher harmonics in Section~\ref{sec:HH}, and, finally, we summarize our results in Section~\ref{sec:conclusions}.

%

\section{The model} \label{sec:model}


We consider an extension of the swamalator model introduced by O'Keeffe \textit{et al.} \cite{OKeeffeEtAl2017} to include second harmonic coupling in the phase dynamics. The spatial $\bm{x}$ and phase $\phi$ dynamics of the $i$-th swarmalator are given by
\begin{align}
\dot{\bm{x}}_i &= \frac{1}{N} \sum_{j=1, \, j\neq i}^N \frac{\bm{x}_j - \bm{x}_i}{|\bm{x}_j - \bm{x}_i|} \left( 1 + J \cos(\phi_j - \phi_i) \right) -  \frac{\bm{x}_j - \bm{x}_i}{|\bm{x}_j - \bm{x}_i|^2}, \label{eq:swarmalator_space} \\
\dot{\phi}_i &= \frac{1}{N} \sum_{j=1, \, j\neq i}^N \frac{1}{|\bm{x}_j - \bm{x}_i|} \left( K_1 \sin(\phi_j - \phi_i) + K_2 \sin\left( 2 (\phi_j - \phi_i ) \right) \right), \label{eq:swarmalator_phase}
\end{align}
where $N$ is the number of swarmalators, $-1\leq J\leq 1$ is a parameter that controls the effect of phase alignment on spatial attraction, and $K_1$ and $K_2$ are phase coupling strengths for the first and second harmonic, respectively. 
We note that the original model is recovered by setting $K_1 = K$ and $K_2 = 0$. Since phase variables are $2\pi$-periodic and coupling is generally anti-symmetric (equal and opposite), general phase coupling functions are $2\pi$-periodic and odd. Hence, general coupling functions can be expressed as a Fourier sine series. The phase dynamics (\ref{eq:swarmalator_phase}) represents a truncated Fourier sine series of more general coupling functions.  

For $K_2>0$, the second harmonic phase coupling creates phase attraction for phase differences close to both $0$ and $\pi$, rather than just $0$ as in the original model. As such, a common feature in the second-harmonic swarmalator model (\ref{eq:swarmalator_space})-(\ref{eq:swarmalator_phase}) is the occurrence of clusters of swarmalators, with the clusters having phases offset by $\pi$. We note that this is also exhibited by Kuramoto-like phase oscillators with second harmonic interaction \cite{HanselEtAl1993, Daido1996, SkardalEtAl2011, BernerEtAl2023}. As we shall see, in some cases there is both phase and spatial clustering, such that two distinct groups emerge, separated both in space and in phase, in other cases there is only phase clustering, such that the swarmalators form a single spatial cluster but have two distinct phases. There are also many interesting non-stationary phenomena observed, including rotating clusters that are fixed in space, and cases such that swarmalators will spend a long period of time in one cluster then rapidly switch phase and move to the other cluster. Here we focus primarily on the static clustered states.

\section{Static single phase cluster} \label{sec:static_single_cluster}

\begin{figure}[tbp]
\centering
\includegraphics[width=0.5\columnwidth]{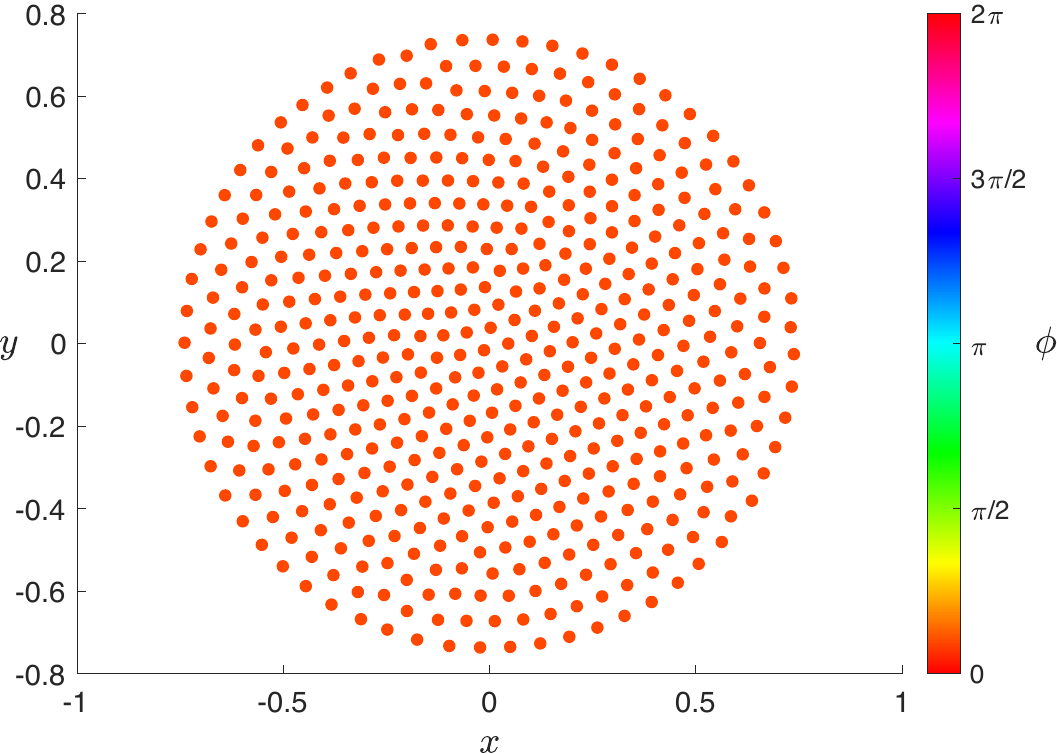}
\caption{
Static single cluster state with all swarmalators having identical phases. The swarmalator model (\ref{eq:swarmalator_space})-(\ref{eq:swarmalator_phase}) parameters are $J=0.5$, $K_1 = 1$, $K_2=0$ and $N=500$.
}
\label{fig:single_cluster_single_phase}
\end{figure}

As a means to better understand clustered states, we begin by discussing the stability of a single static spatial cluster of swarmalators, with all swarmalators having identical phases. An example of such a state is shown in Fig.~\ref{fig:single_cluster_single_phase}. These states are found to be stable in the original model \cite{OKeeffeEtAl2017} ($K_2=0$) for all values of $J$ provided $K_1>0$. In this section we generalize this stability result when second-harmonic coupling is included ($K_2 \neq 0$).

For the static single cluster state, we determine asymptotic stability of the full dynamics (\ref{eq:swarmalator_space})-(\ref{eq:swarmalator_phase}) by first considering the linear stability of the purely phase dynamics (\ref{eq:swarmalator_phase}) for an arbitrary stationary spatial configuration. If the state with identical phases is stable for all stationary spatial configurations, then it is also stable when the spatial configuration is time-dependent, i.e., in the full model. We can then conclude that the static single cluster state is stable in the full model. This follows from the fact that the spatial dynamics with constant phases can be written as a gradient system
\begin{equation} \label{eq:space_potential}
\dot {\bm{x}}_i = \frac{1}{N} \sum_{j=1, \, j \neq i}^N - \nabla U_{ij}\left( \bm{x}_i - \bm{x}_j \right)
\end{equation}
where $U_{ij}(\bm{x}) = | \bm{x} | \left(1+J \cos(\phi_j - \phi_i) \right) - \log |\bm{x}|$ is the interaction potential, with $U_{ij} = U_{ji}$. As such, for any stationary set of phases the spatial dynamics converges to a stable stationary state.

Conversely, if the identical phase state is unstable for all stationary spatial configurations, then it is clear that there can be no stable stationary state with all swarmalators having identical phase.

\begin{figure}[tbp]
\centering
\includegraphics[width=0.7\columnwidth]{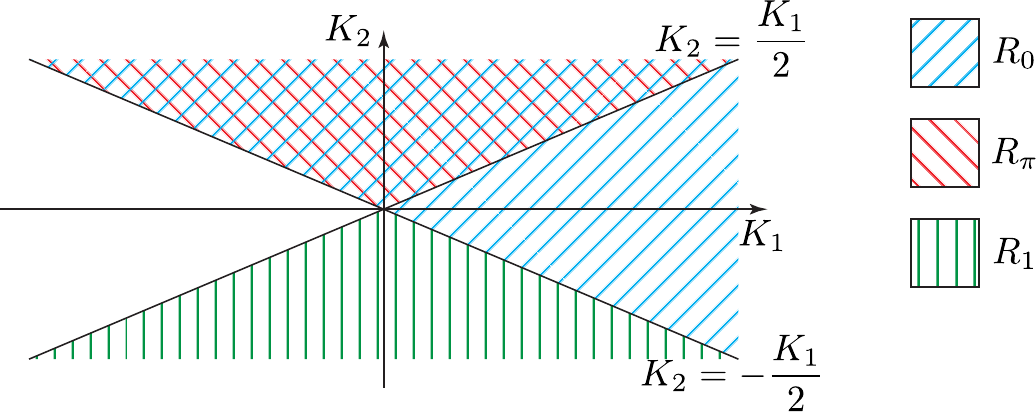}
\caption{
Stability regions $R_0$ (\ref{eq:stability_R0}) (blue), $R_\pi$ (\ref{eq:stability_Rpi}) (red) and $R_1$ (\ref{eq:stability_R1}) (green).
}
\label{fig:stability_regions}
\end{figure}

Consider a fixed static spatial configuration $\bm{x}_i$ of the swarmalators, and the purely phase dynamics given by (\ref{eq:swarmalator_phase}). The Jacobian of the phase dynamics is given by
\begin{equation}
\left(\mathcal{J}(\bm{\phi})\right)_{ij} = \frac{\partial \dot\phi_i}{\partial \phi_j}  = 
\frac{1}{N} 
\begin{cases} 
- \sum_{k \neq i} \frac{K_1 \cos(\phi_k - \phi_i) + 2 K_2 \cos(2(\phi_k - \phi_i))}{|\bm{x}_k - \bm{x}_i|}, & i=j, \\
\frac{K_1 \cos(\phi_j - \phi_i) + 2 K_2 \cos(2(\phi_j - \phi_i))}{|\bm{x}_j - \bm{x}_i|}, & i \neq j.
\end{cases}
\end{equation}
We note that $\mathcal{J}$ always has an eigenvalue $\lambda = 0$ with eigenvector $(1,1,\dots,1)$, which corresponds to the invariance of the system to constant phase shifts.  For the state with all phases identical, $\phi_i = \phi^*$ for all $i$, the Jacobian is equal to
\begin{align}
\left(\mathcal{J}(\bm{\phi}^*)\right)_{ij}  &= 
-\frac{K_1 + 2 K_2}{N} 
\begin{cases} 
 \sum_{k \neq i} \frac{1 }{|\bm{x}_k - \bm{x}_i|}, & i=j,  \\
-\frac{1 }{|\bm{x}_j - \bm{x}_i|}, & i \neq j,
\end{cases} \nonumber \\
&= - \frac{K_1 + 2 K_2}{N}  \mathcal{L}_{ij},  \label{eq:single_cluster_Jacobian}
\end{align}
where $\mathcal{L}$ is the graph Laplacian of the weighted undirected network with adjacency matrix $A_{ij} = \frac{1 }{|\bm{x}_j - \bm{x}_i|}$. Graph Laplacians are positive semi-definite, with nullity equal to the number of connected components. For the graph Laplacian $\mathcal{L}$ here, the graph is fully connected, and so the zero eigenvalue has multiplicity equal to one.  From (\ref{eq:single_cluster_Jacobian}) it follows that the spectrum of $\mathcal{J}(\bm{\phi}^*)$ can be split into three cases:
\begin{enumerate}[(i)]
	\item $K_1 + 2 K_2 > 0$: $\lambda_1 = 0$ and $\lambda_i < 0$ for $i\geq 2$, and, hence, the identical phase state is asymptotically stable.
	\item $K_1 + 2K_2 < 0$: $\lambda_1 = 0$ and $\lambda_i >0$ for $i \geq 2$, and, hence, the identical phase state is unstable.
	\item $K_1 + 2K_2 = 0$: $\lambda_i = 0$ for all $i$. Stability cannot be inferred.
\end{enumerate}
Therefore, a perturbation in phases away from an identical state will decay for all fixed spatial configurations provided that the parameters are in the region
\begin{equation} \label{eq:stability_R0}
R_0 = \left\lbrace (J,K_1,K_2): K_2 > -K_1/2 \right \rbrace.
\end{equation}
Hence, the static single cluster state is stable in the full system (\ref{eq:swarmalator_space})-(\ref{eq:swarmalator_phase}) for parameters in the region $R_0$, which is shaded blue in Fig.~\ref{fig:stability_regions}.

We note that in the parameter region $K_1<0$ and $K_2<0$, i.e., the third quadrant of Fig.~\ref{fig:stability_regions}, the dynamical regimes can mostly be categorized by those already found in the original swarmalator model \cite{OKeeffeEtAl2017}, i.e., static async and phase waves. Here we focus primarily on the novel clustered states that arise due to the presence of the second harmonic interaction.


\section{States with two distinct phases}  \label{sec:2_phase_stability}

We apply a similar reasoning to consider the stability of stationary states for which the phases take on exactly two values, i.e., $\phi_i = \theta_1$ for $i \in \mathcal{C}_1$ and $\phi_i = \theta_2$ for $i \in \mathcal{C}_2$. These states are expected due to the inclusion of the second harmonic in the phase dynamics (\ref{eq:swarmalator_phase}). We again focus on the purely phase dynamics for static spatial configurations, and determine sufficient conditions for the system parameters and phases $\theta_{1,2}$ for which these states are stationary and stable. As in Section~\ref{sec:static_single_cluster}, if the phase dynamics are stable, then it follows that there exists a stable stationary spatial state corresponding to that set of phases, i.e., a stable stationary solution of the full system (\ref{eq:swarmalator_space})-(\ref{eq:swarmalator_phase}).

Assuming all phases take on one of two values, $\theta_1$ or $\theta_2$, the phase dynamics (\ref{eq:swarmalator_phase}) are stationary if and only if
\begin{equation}
0=K_1 \sin\Phi + K_2 \sin 2\Phi = \sin \Phi \left( K_1 + 2K_2 \cos\Phi \right),
\end{equation}
where $\Phi = \theta_2 - \theta_1$. Solutions satisfy one of three cases:

	\noindent \textit{Case 1:} $\Phi = 0$, i.e., all swarmalators have identical phase, reducing to the static single phase cluster case in Section~\ref{sec:static_single_cluster}.
	
	\noindent \textit{Case 2:} $\Phi = \pi$, corresponding to anti-phase sets of swarmalators.
	
	\noindent \textit{Case 3:} $\cos \Phi = -\frac{K_1}{2K_2}$ with $\Phi \neq 0,\pi$.

In all these cases, the Jacobian of the purely phase dynamics is given by
\begin{equation}
\left(\mathcal{J}(\bm{\phi})\right)_{ij} = \frac{1}{N} \begin{cases}
-\left(\sum_{k\in \mathcal{C}_m, k\neq i} \frac{K_1 + 2K_2}{|\bm{x}_k - \bm{x}_i|} \right)
-\left(\sum_{k\notin \mathcal{C}_m,} \frac{K_1 \cos \Phi + 2K_2 \cos 2\Phi}{|\bm{x}_k - \bm{x}_i|} \right) , & j=i, \\
 \frac{K_1 + 2K_2}{|\bm{x}_j - \bm{x}_i|},  &j \in \mathcal{C}_m, \, j\neq i ,  \\
 \frac{K_1 \cos \Phi + 2K_2 \cos 2\Phi}{|\bm{x}_j - \bm{x}_i|}, & j\notin \mathcal{C}_m,
\end{cases}
\end{equation}
for $i \in \mathcal{C}_m$ and $m=1,2$. This Jacobian is again closely related to a graph Laplacian. Explicitly, $\mathcal{J} = -\mathcal{L}/N = -(D-A)/N$ where the adjacency matrix $A$ is equal to
\begin{equation}
A_{ij} = \begin{cases}
0, & i=j, \\
 \frac{K_1 + 2K_2}{|\bm{x}_j - \bm{x}_i|}, & i,j \in \mathcal{C}_m, \, i\neq j ,\\
  \frac{K_1 \cos \Phi + 2K_2 \cos 2\Phi}{|\bm{x}_j - \bm{x}_i|}, & i\in \mathcal{C}_m, \, j\in \mathcal{C}_n \text{ with } m\neq n.
\end{cases}
\end{equation}
This adjacency matrix corresponds to a weighted undirected graph, but may have negative edge weights. In cases where all edge weights are positive, the graph Laplacian $\mathcal{L}$ is positive semi-definite, with a single zero eigenvalue, and so the Jacobian $\mathcal{J}$ has all negative eigenvalues except the single zero eigenvalue corresponding to phase-shift invariance. 

For Case 2, i.e., $\Phi=\pi$ and the swarmalators are anti-phase, all edge weights of $A$ are positive if and only if $K_1 + 2K_2 >0$ and $-K_1 + 2K_2>0$. Therefore, in the region 
\begin{equation} \label{eq:stability_Rpi}
R_\pi = \left\lbrace(J, K_1, K_2) :\, K_2 > -K_1/2 \text{ and } K_2 > K_1/2\right\rbrace ,
\end{equation}
of parameter space there exist anti-phase stable stationary solutions of the full dynamics (\ref{eq:swarmalator_space})-(\ref{eq:swarmalator_phase}). The region $R_\pi$ is shaded red in Fig.~\ref{fig:stability_regions}.

\begin{figure}[tbp]
\centering
\includegraphics[width=\columnwidth]{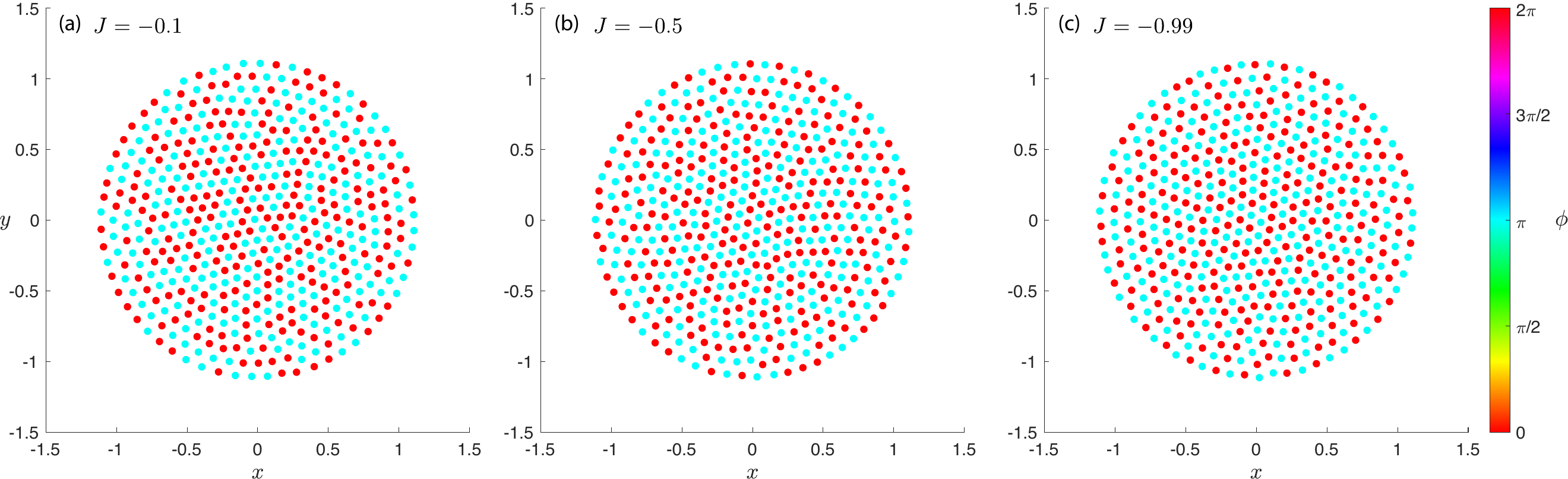}
\caption{
Static anti-phase single cluster states for (a)~$J=-0.1$, (b)~$J=-0.5$, and (c)~$J=-0.99$. The swarmalator model (\ref{eq:swarmalator_space})-(\ref{eq:swarmalator_phase}) parameters for all are $N=500$, $K_1 = -0.5$ and $K_2 = 0.5$.
}
\label{fig:2_phase_1_cluster}
\end{figure}

For $J<0$, such that opposites attract, the anti-phase states form a single spatial cluster with a crystalline lattice structure, as shown in Fig.~\ref{fig:2_phase_1_cluster}. One can see that for $J\approx 0$ (e.g., Fig.~\ref{fig:2_phase_1_cluster}(a)), there are several small clusters of swarmalators with the same phase, whereas for $J\approx 1$ (e.g., Fig.~\ref{fig:2_phase_1_cluster}(c)), the phases are more well mixed. This will be studied in more detail in Section~\ref{sec:anti-phase_one_cluster}.

\begin{figure}[tbp]
\centering
\includegraphics[width=\columnwidth]{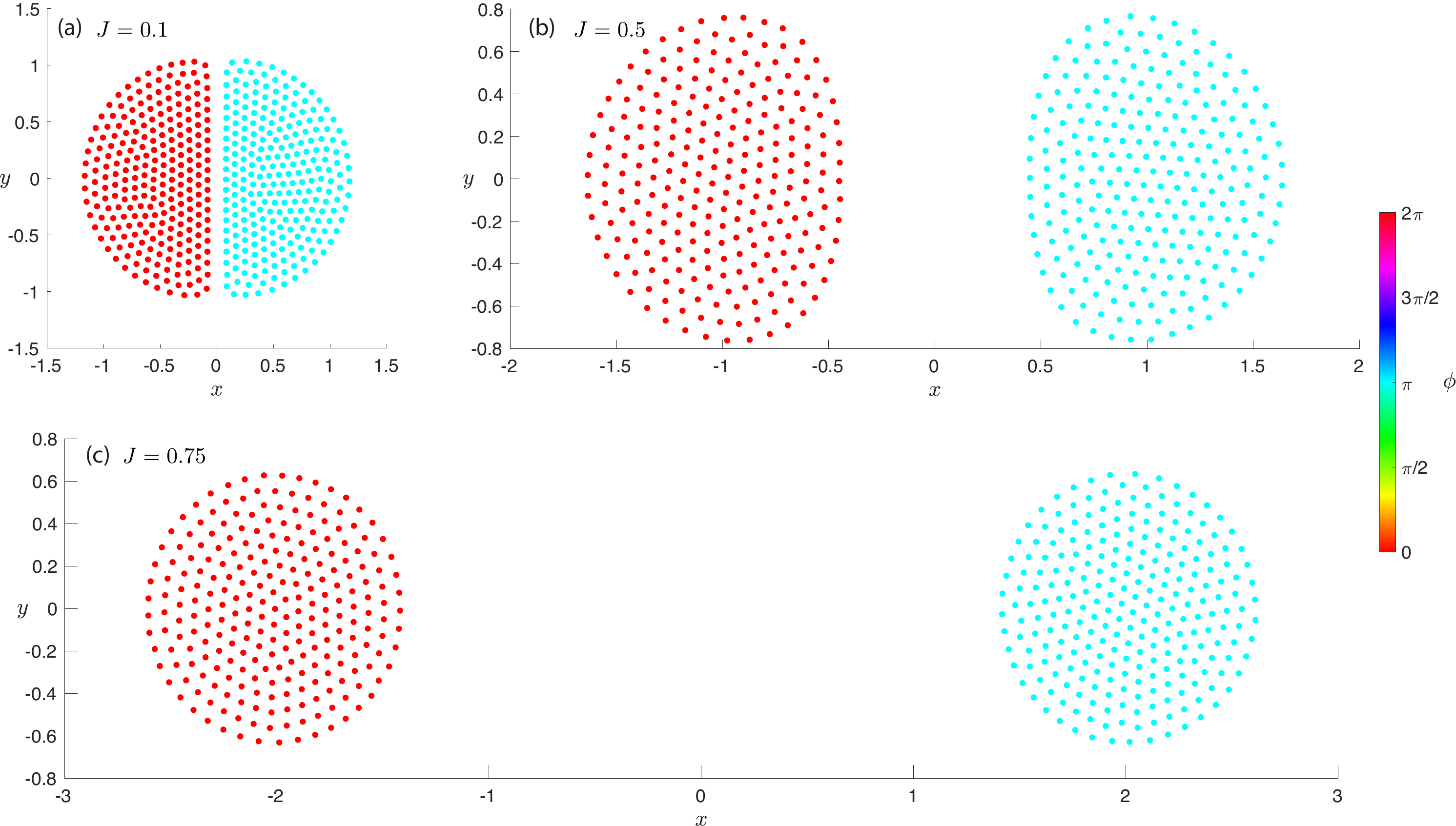}
\caption{
States with two static anti-phase clusters using (a) $J=0.1$, (b) $J=0.5$ and (c) $J=0.75$. The swarmalator model (\ref{eq:swarmalator_space})-(\ref{eq:swarmalator_phase}) parameters for all are $N=500$, $K_1 = -0.5$ and $K_2 = 0.5$.
}
\label{fig:2_phase_2_cluster}
\end{figure}

For $J>0$, such that like-attracts-like,  the anti-phase states arrange themselves into two distinct spatial clusters, as shown in Fig.~\ref{fig:2_phase_2_cluster}, with the distance between the clusters increasing as $J$ increases. In Section~\ref{sec:anti-phase_clusters} a mean-field approximation will be used to derive an analytic approximation for the cluster separation distance as a function of $J$.

\begin{figure}[tbp]
\centering
\includegraphics[width=0.4\columnwidth]{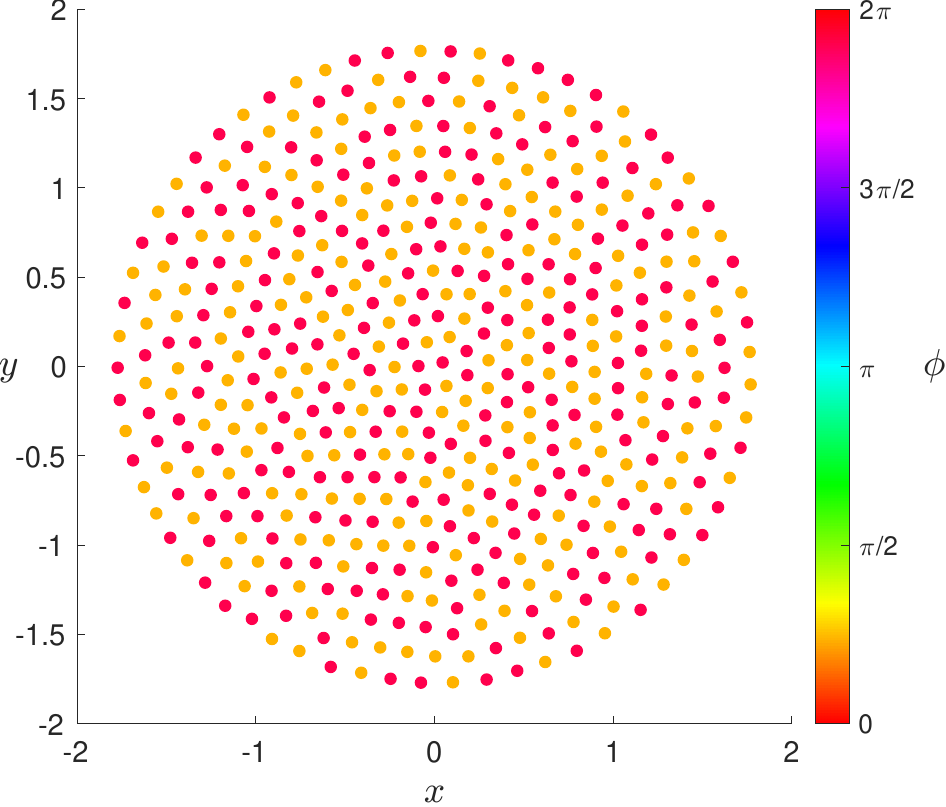}
\caption{
Static single cluster state with two distinct phases $\phi_1 = 5.9875$ and $\phi_2 = 0.7515$ for the swarmalator model (\ref{eq:swarmalator_space})-(\ref{eq:swarmalator_phase}) parameters $J=-0.5$, $K_1 = 0.5$, $K_2 = -0.5$ and $N=500$.
}
\label{fig:non-anti-phase_cluster}
\end{figure}

For Case 3, in which there are two distinct phases, but they are not anti-phase, instead having $\cos\Phi = -\frac{K_1}{2K_2}$ with $\Phi \neq 0,\pi$, we first note solutions for $\Phi$ only exist for parameters in the regions $R_\pi$ (\ref{eq:stability_Rpi}) and
\begin{equation} \label{eq:stability_R1}
R_1 =  \{(J, K_1, K_2) :\, K_2 < -K_1/2 \text{ and } K_2 < K_1/2\} .
\end{equation}
The region $R_1$ is shaded green in Fig.~\ref{fig:stability_regions}. All edge weights of $A$ are positive if and only if $K_1 + 2K_2 >0$ and $K_1 \cos\Phi + 2K_2 \cos 2\Phi >0$.  The first inequality can only be true if the parameters belong to $R_\pi$. Substituting $\cos\Phi = -\frac{K_1}{2K_2}$, the second inequality corresponds to the region
\begin{equation}
\frac{K_1^2}{2K_2} - 2K_2 >0,
\end{equation}
which can only be true if the parameters belong to $R_1$. Therefore, in order for both inequalities to be satisfied, the parameters must belong to both $R_\pi$ and $R_1$, an impossibility since these sets are disjoint (cf. Fig.~\ref{fig:stability_regions}). This means that there is no region of parameter space for which all edge weights of $A$ are positive. This does not rule out the possibility of two-phase states satisfying $\cos\Phi = -\frac{K_1}{2K_2}$. There exist stable stationary states of the full dynamics (\ref{eq:swarmalator_space})-(\ref{eq:swarmalator_phase}) with the adjacency matrix $A$ having some negative edge weights. An example is shown in Fig.~\ref{fig:non-anti-phase_cluster} using the parameters $J=-0.5$, $K_1 = 0.5$, $K_2 = -0.5$ and $N=500$. In this example the swarmalators form a static single cluster with two distinct phases, $\phi_1 = 5.9875$ and $\phi_2 = 0.7515$, yielding a phase difference $\Phi = 1.0472$ (using $\phi_1 = -0.2957$) which agrees with $\cos\Phi = -\frac{K_1}{2K_2}$.

\section{Anti-phase single cluster} \label{sec:anti-phase_one_cluster}


Having found in Section~\ref{sec:2_phase_stability} that anti-phase static states are stable in the region $R_\pi$, we now study properties of the stationary states that emerge from random initial conditions in the cases where $J<0$ and $J>0$. In this section we focus on the states that arise when $J<0$, examples of which are shown in Fig.~\ref{fig:2_phase_1_cluster}. These states can be described as having a single spatial cluster with opposite phases. In Section~\ref{sec:anti-phase_clusters} we will explore the case $J>0$.

Since we are considering anti-phase states, and the system (\ref{eq:swarmalator_space})-(\ref{eq:swarmalator_phase}) is invariant to uniform phase shifts of all swarmalators, we may assume, without loss of generality, that all swarmalators in the long-term have phase either $0$ or $\pi$. As examples, the phases in Fig.~\ref{fig:2_phase_1_cluster} and Fig.~\ref{fig:2_phase_2_cluster} have been uniformly shifted to this effect.

\begin{figure}[tbp]
\centering
\includegraphics[width=\columnwidth]{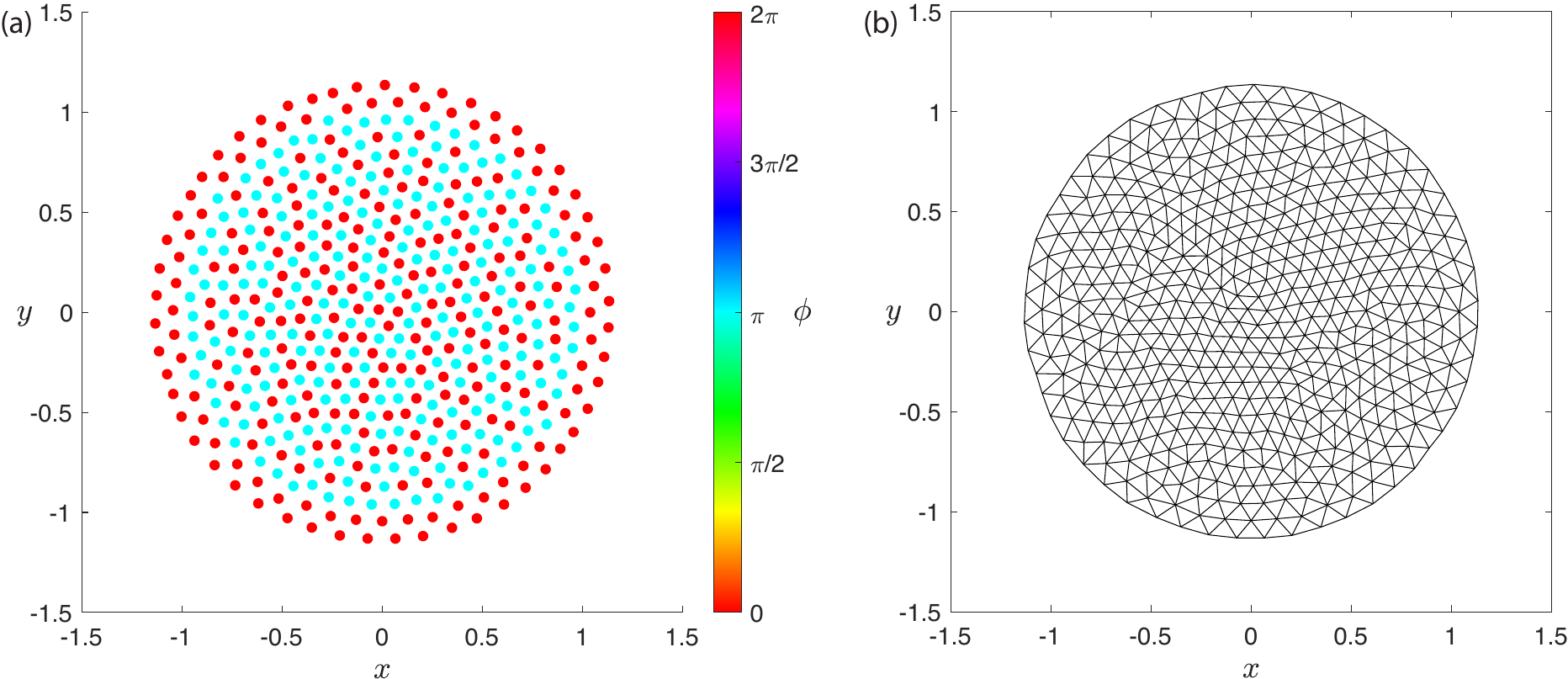}
\caption{
(a)~Single cluster anti-phase state from the swarmalator model (\ref{eq:swarmalator_space})-(\ref{eq:swarmalator_phase}) with $J=-0.99$, $K_1=-0.5$, $K_2 = 0.5$ and $N=500$. (b)~Corresponding Delaunay triangulation of the swarmalator positions.
}
\label{fig:2_phase_1_cluster_outer_ring_and_tri}
\end{figure}

\begin{figure}[tbp]
\centering
\includegraphics[width=0.5\columnwidth]{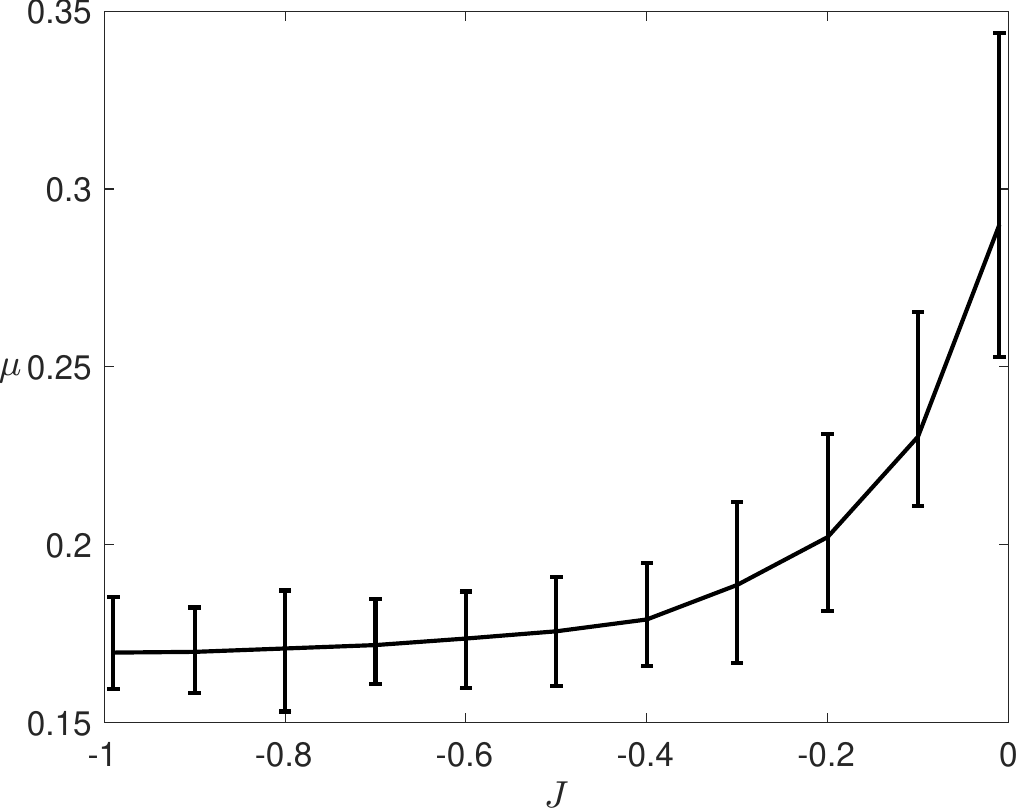}
\caption{
The mixing metric $\mu$ as $J$ is varied. The mean, maximum and minimum values of $\mu$ are shown for 100 random initial conditions at each value of $J$. The swarmalator model (\ref{eq:swarmalator_space})-(\ref{eq:swarmalator_phase}) parameters $K_1=-0.5$, $K_2 = 0.5$ and $N=500$ are used for all simulations.
}
\label{fig:2_phase_1_cluster_mixing}
\end{figure}

As discussed briefly in the previous section, as $J$ decreases (becoming more negative), the mixing between the $0$-phase and $\pi$-phase swarmalators increases. For $J=-0.1$ (Fig.~\ref{fig:2_phase_1_cluster}(a)) there are several small clusters of same-phase swarmalators, whereas no such clustering is evident for $J=-0.99$ (Fig.~\ref{fig:2_phase_1_cluster}(c)). Instead, at $J=-0.99$ there are alternating ``stripes'' of $0$-phase  and $\pi$-phase swarmalators. We introduce a mixing metric which utilizes the local phase order parameter of each of the swarmalators. In a cluster of same-phase swarmalators, the local order parameter for each swarmalator will be close to one, and so in a poorly mixed state such as Fig.~\ref{fig:2_phase_1_cluster}(a) with many same-phase clusters, many of the local order parameters will be close to one, and the average local order parameter (averaging over all swarmalators) will be close to one. Conversely, in a well-mixed state such as Fig.~\ref{fig:2_phase_1_cluster}(c), the local order parameters will be close to zero, since each swarmalator is surrounded by approximately the same number of same-phase and anti-phase swarmalators. Hence, in a well-mixed state the average local order parameter will be close to zero. To define the local order parameter, we define connectedness of swamalators using a Delaunay triangulation, which yields a triangulation adjacency matrix $T$. An example of such a triangulation is shown in Fig.~\ref{fig:2_phase_1_cluster_outer_ring_and_tri}(b) corresponding to the stationary state of the system shown in Fig.~\ref{fig:2_phase_1_cluster_outer_ring_and_tri}(a). For each swarmalator $j$, the local order parameter $r_j$ is defined as
\begin{equation} \nonumber
r_j = \left| \frac{1}{d_j+1} \left( e^{i \phi_j} +  \sum_{k: T_{jk} = 1}  e^{i \phi_k} \right) \right|,
\end{equation}
where $d_j$ is the degree of swarmalator $j$ in the Delaunay triangulation. The local order parameter is the order parameter of all nodes connected to $j$. We note that when there is an imbalance between the number of $0$-phase and $\pi$-phase swarmalators, the larger population will form a ring around a mixed interior, as demonstrated in Fig.~\ref{fig:2_phase_1_cluster_outer_ring_and_tri}(a) where a ring of $0$-phase swarmalators encloses a well-mixed interior. Therefore, when defining the average local order parameter, we average only over interior nodes, i.e., those satisfying $|\bm{x}_j| < 0.7R$, where $R = \max\{ |\bm{x}_j - \bar{\bm{x}}|\}$ is the radius of the cluster. The average local order parameter, which quantifies the degree of mixing in the interior of the cluster, is defined as
\begin{equation}
\mu = \frac{1}{N_{0.7R}} \sum_{|\bm{x}_j| < 0.7R} r_j,
\end{equation}
where $N_{0.7R}$ is the number of swarmalators in the interior. Fig.~\ref{fig:2_phase_1_cluster_mixing} shows that the mixing metric $\mu$ decreases as $J$ decreases (becoming more negative). This confirms that the degree of mixing between the $0$-phase and $\pi$-phase swarmalators increases as $J$ decreases, as suggested visually by Fig.~\ref{fig:2_phase_1_cluster}.

The Delaunay triangulation reveals an approximately hexagonal crystalline structure with some imperfections. As such, most swarmalators have six neighbors in the triangulation. For any swarmalator with six neighbors, the local order parameter $r_j$ averages over 7 swarmalators, and, thus, can never be zero in an anti-phase state. The minimum of $r_j$ for a swarmalator with six neighbors in an anti-phase state is $r_j = 1/7 \approx 0.143$. As such, a soft lower bound on the mixing metric $\mu$ is $1/7$ \footnote{Imperfections in the lattice which yield swarmalators with 5 or 7 neighbors can have $r_j =0$, so $1/7$ is not a strict lower bound.}. Fig.~\ref{fig:2_phase_1_cluster_mixing} shows that as $J$ approaches $-1$, the mixing metric $\mu$ approaches this lower bound, confirming that nearly optimal mixing is achieved.

\begin{figure}[tbp]
\centering
\includegraphics[width=0.5\columnwidth]{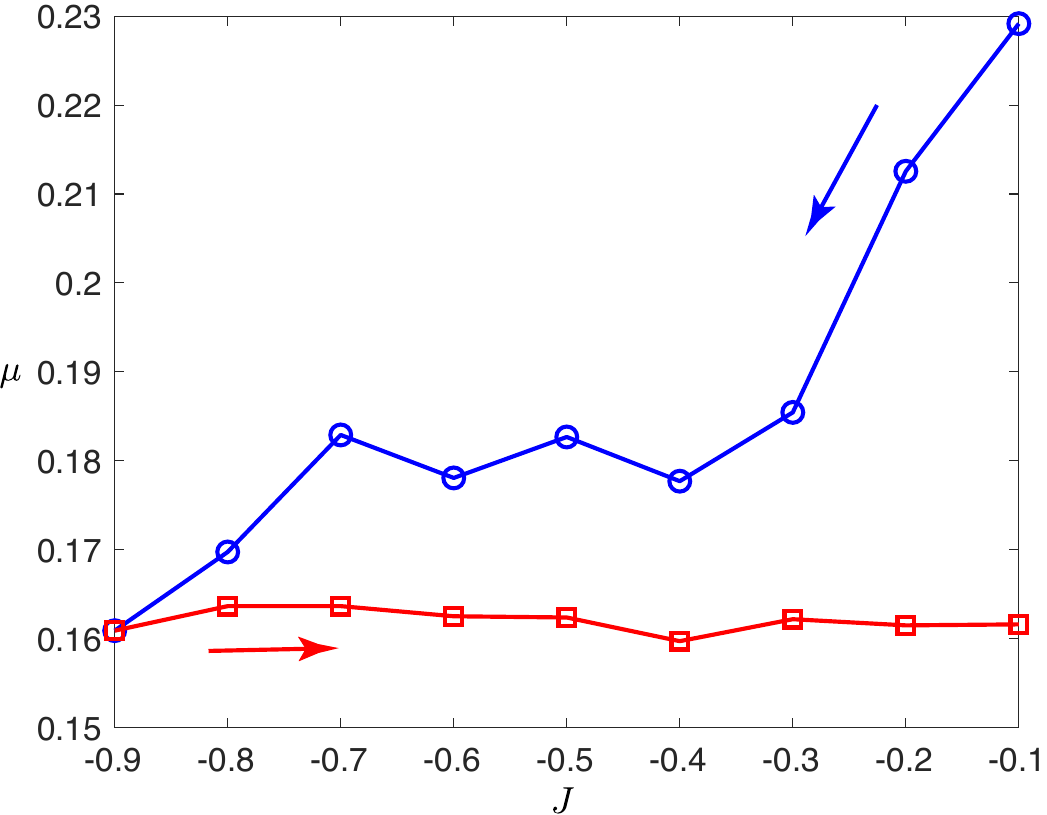}
\caption{
The mixing metric $\mu$ as $J$ is varied. Started at $J=-0.1$ from a random IC until equilibrium is reached. The equilibrium state from the $J=-0.1$ simulation is used as the initial condition for decreasing values of $J$ (blue circles). The equilibrium state from $J=-0.9$ is then used as the initial condition for increasing values of $J$ (red squares).
}
\label{fig:2_phase_1_cluster_end_stability_mixing}
\end{figure}

We remark that while random initial conditions generally do not lead to well-mixed equilibrium solutions for $J\approx 0$, there do exist well-mixed equilibrium states. We show this by testing whether a well-mixed state will un-mix as $J$ is increased. We start with a random initial condition and simulate the system (\ref{eq:swarmalator_space})-(\ref{eq:swarmalator_phase}) until equilibrium is reached for $J=-0.1$. The resulting equilibrium is the poorly mixed state shown in Fig.~\ref{fig:2_phase_1_cluster}(a). We then use the $J=-0.1$ equilibrium state as the initial condition for decreasing values of $J$ and compute the mixing metric $\mu$. As expected, the mixing increases ($\mu$ decreases) as $J$ decreases, as shown in Fig.~\ref{fig:2_phase_1_cluster_end_stability_mixing} by the blue circles. Next, to test whether a well-mixed state will un-mix, we reverse the process. We use the equilibrium found for $J=-0.9$ as the initial condition for increasing values of $J$ and again compute the mixing metric $\mu$. The results are shown in Fig.~\ref{fig:2_phase_1_cluster_end_stability_mixing} by the red squares. It is found that the mixed state does not un-mix. This means that there exist stable well-mixed equilibrium states for all values of $J$, but Fig.~\ref{fig:2_phase_1_cluster_mixing} shows that poorly mixed states occur more frequently for random initial conditions and $J\approx 0$.

Fig.~\ref{fig:2_phase_1_cluster_end_stability_mixing} also shows hysteresis in the mixing and un-mixing process. A decrease in $J$ followed by and increase in $J$ will yield a new, better mixed equilibrium solution.


As well as measuring the degree of mixing between the anti-phase groups, we also measured the size of the cluster. It is found that the size increases only slightly as $J$ decreases. The difference in the mean cluster size between $J=-0.01$ and $J=-0.99$ is $0.5\%$.

\section{Two anti-phase spatial clusters}  \label{sec:anti-phase_clusters}

We now consider the two cluster states that arise in the parameter region $R_\pi$ with $J>0$. We employ a mean-field (centroid) approach to determine the stable separation distance between the clusters. A similar has been considered in \cite{SarEtAl2022}, though for fundamentally different underlying dynamics to that considered here.


Consider the state such that there are two distinct anti-phase clusters. In the first cluster there are $N_1$ swarmalators with indices belonging to $\mathcal{C}_1$, and phases $\phi_i = \theta_1$. In the second cluster there are $N_2 = N-N_1$ swarmalators with indices belonging to $\mathcal{C}_2$, and phases $\phi_i = \theta_1 + \pi$. It is assumed that $N_1$, $N_2$, $\mathcal{C}_1$ and $\mathcal{C}_2$ are known \textit{a priori}. The centroid of each cluster is given by the mean position
\begin{equation}
\bm{X}_k = \frac{1}{N_k} \sum_{i\in \mathcal{C}_k} \bm{x}_i
\end{equation}
for $k=1,2$. The dynamics of the centroids is then obtained by averaging the spatial dynamics (\ref{eq:swarmalator_space}), yielding
\begin{align}
\dot{\bm{X}_1} &= \frac{1}{N N_1} \sum_{i\in \mathcal{C}_1} \sum_{j \neq i} \frac{\bm{x}_j - \bm{x}_i}{|\bm{x}_j - \bm{x}_i|} \left( 1 + J \cos(\phi_j - \phi_i) \right) -  \frac{\bm{x}_j - \bm{x}_i}{|\bm{x}_j - \bm{x}_i|^2}    \\
&= \frac{1}{N N_1}  \left[  \left(\sum_{i\in \mathcal{C}_1} \sum_{j\in \mathcal{C}_1,\, j\neq i} \frac{\bm{x}_j - \bm{x}_i}{|\bm{x}_j - \bm{x}_i|} \left( 1 + J \right) -  \frac{\bm{x}_j - \bm{x}_i}{|\bm{x}_j - \bm{x}_i|^2} \right) + \right. \\ 
&  \left.\left(\sum_{i\in \mathcal{C}_1} \sum_{j\in \mathcal{C}_2}  \frac{\bm{x}_j - \bm{x}_i}{|\bm{x}_j - \bm{x}_i|} \left( 1 - J \right) -  \frac{\bm{x}_j - \bm{x}_i}{|\bm{x}_j - \bm{x}_i|^2} \right) \right],
\end{align}
for the dynamics of $\bm{X}_1$, and similarly for $\bm{X}_2$. We note that due to anti-symmetry in $i$ and $j$, the double sum in the first bracket is zero. For the second double sum we make the approximation that every swarmalator can be identified with the centroid of their respective cluster, i.e., $\bm{x}_i \approx \bm{X}_k$ for all $i \in \mathcal{C}_k$. In addition, by translating and rotating the reference frame we may assume, without loss of generality, that the $y$-components of the centroids $\bm{X}_{1,2}$ are zero, and so we can write $\bm{X}_k = (X_k,0)$, with $X_2 - X_1 =  |\bm{X}_2 - \bm{X}_1| =  R$. With this approximation we obtain
\begin{align}
\dot{X}_1 &= \frac{1}{N N_1} \left( \sum_{i\in \mathcal{C}_1} \sum_{j\in \mathcal{C}_2}  \frac{X_2 - X_1}{|X_2 - X_1|} \left( 1 - J \right) -  \frac{X_2 - X_1}{|X_2 - X_1|^2} \right)  \nonumber \\
&=  \frac{ N_2}{N } \left( 1 - J - \frac{1}{R} \right),  \label{eq:two_cluster_position_1} \\
\dot{X}_2 &=  \frac{ N_1}{N } \left( -(1 - J) + \frac{1}{R} \right). \label{eq:two_cluster_position_2}
\end{align}
Taking the difference of (\ref{eq:two_cluster_position_1}) and (\ref{eq:two_cluster_position_2}) yields the evolution equation of the cluster separation $R$
\begin{equation} \label{eq:two_cluster_R}
\dot{R} = \dot{X}_2 - \dot{X}_1 = J-1 + \frac{1}{R}.
\end{equation}
Interestingly, the cluster separation dynamics do not depend on the absolute or relative sizes of the clusters. While random initial conditions generally yield approximately equal sized clusters, states with $N_1 = 1$ and $N_2 = N-1$ are stable for all values of $N$.

\begin{figure}[tbp]
\centering
\includegraphics[width=0.6\columnwidth]{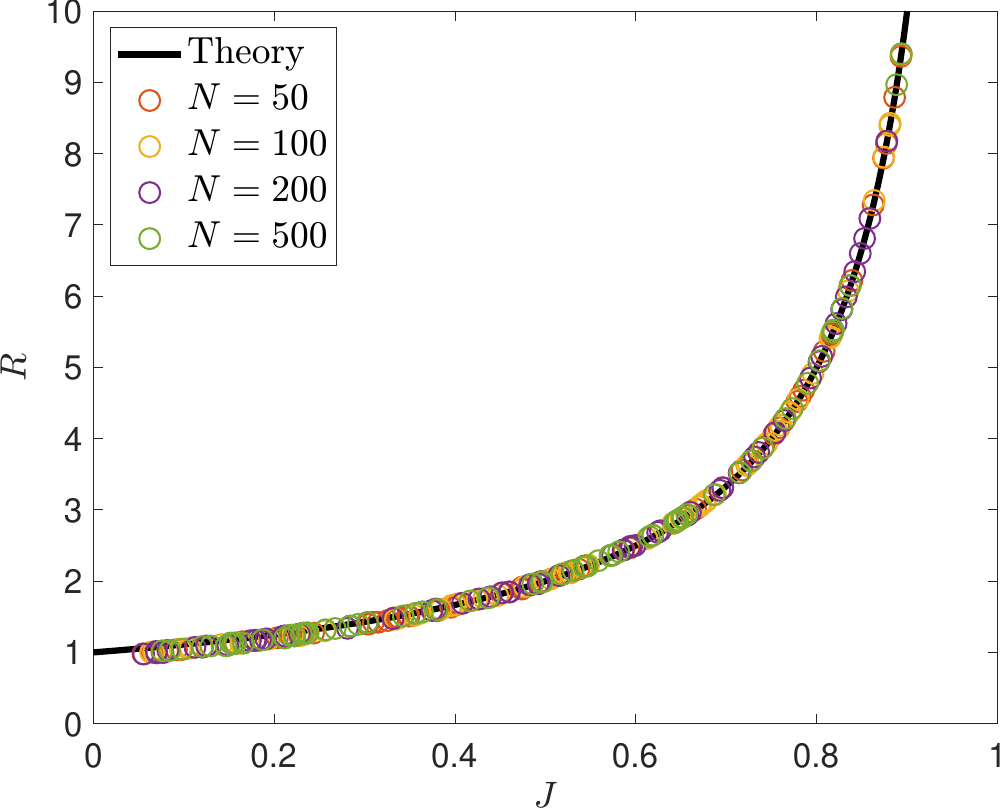}
\caption{
Cluster separation distance $R$ for different values of $0<J<1$. The theoretical value (\ref{eq:two_cluster_sep_distance}) (black curve) is shown together with realizations of the full model (\ref{eq:swarmalator_space})-(\ref{eq:swarmalator_phase}) (colored circles). For each of $N=50,\,100,\,200,\,500$, the full model is simulated using 64 random realizations of the parameters $J$, $K_1$, $K_2$ and $N_1/N$.
}
\label{fig:two_cluster_sep_distance}
\end{figure}

The (stable) stationary solution to (\ref{eq:two_cluster_R}) is 
\begin{equation} \label{eq:two_cluster_sep_distance}
R^* = \frac{1}{1-J}.
\end{equation}
Comparing this theoretical approximation with simulations of the full system (\ref{eq:swarmalator_space})-(\ref{eq:swarmalator_phase}), 
Fig.~\ref{fig:two_cluster_sep_distance} shows that there is very good agreement between the theoretical approximation (\ref{eq:two_cluster_sep_distance}) and the computed distance from simulations of the full system for a wide range system parameters $J$, $K_1$, $K_2$, $N$, $N_1$ and $N_2$. For each value of $N=50,\, 100,\, 200,\, 500$, simulations were performed for $64$ random realizations of $J$, $K_1$, $K_2$ and $N_1/N$. We observe that the theoretical approximation (\ref{eq:two_cluster_sep_distance}) is most accurate for $J\approx 1$, which corresponds to large separations $R\gg 1$, while the approximation is least accurate for $J\approx 0$, which corresponds to smaller separation distances. This is because the approximation of the individual swarmalator positions as their cluster centroids is more accurate if the clusters are further apart, such as in Fig.~\ref{fig:2_phase_2_cluster}(c), and in turn the clusters themselves are more circular. Conversely, if the clusters are very close together, as in Fig.~\ref{fig:2_phase_2_cluster}(a), then they become `squashed' together and the centroid approximation is less accurate.

\section{Two static anti-phase clusters and vacillators} \label{sec:vacillators}

\begin{figure}[tbp]
\centering
\includegraphics[width=0.7\columnwidth]{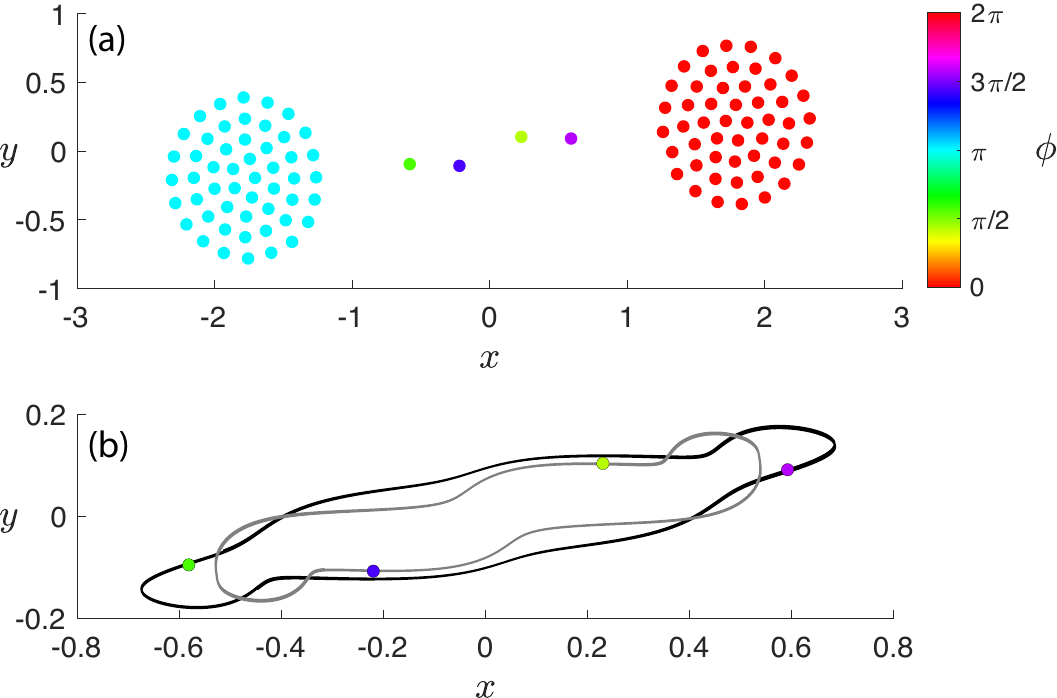}
\caption{
(a)~Four vacillators trapped between two large anti-phase clusters. Simulation from the swarmalator model (\ref{eq:swarmalator_space})-(\ref{eq:swarmalator_phase}) with $J=0.75$, $K_1 = -0.5$, $K_2=0.25$ and $N = 106$. (b)~The trajectories of the four vacillators. Two follow the gray path and two follow the black path.
}
\label{fig:four_vacillator}
\end{figure}

Near the edge of $R_\pi$, with $K_2>0$ and $K_1 \approx -2 K_2$, random initial conditions often converge to a state such that there are two large anti-phase clusters, as predicted by the two cluster model (\ref{eq:two_cluster_R}), but with a small number of swarmalators trapped between the two clusters. These swarmalators typically undergo complex oscillatory dynamics, with both spatial and phase oscillations. We term these trapped swarmalators as ``vacillators'' since they waver between the two anti-phase groups. An example with four vacillators is shown in Fig.~\ref{fig:four_vacillator}. There are two vacillators whose phases stay in the range $(0,\pi)$, and two other vacillators whose phases stay in the range $(\pi,2\pi)$. All four vacillators periodically waver between the two large clusters, following the black and gray paths shown in Fig.~\ref{fig:four_vacillator}(b) (two on each path). In this case the vacillator dynamics is periodic and possesses symmetries, but this is not always true, and we conjecture that irregular chaotic dynamics is possible if there are sufficiently many vacillators. We note that accurate numerical simulation of (\ref{eq:swarmalator_space})-(\ref{eq:swarmalator_phase}) with large $N$ and many vacillators is challenging because the system becomes stiff. This is discussed in more detail later in this section.

\begin{figure}[tbp]
\centering
\includegraphics[width=0.85\columnwidth]{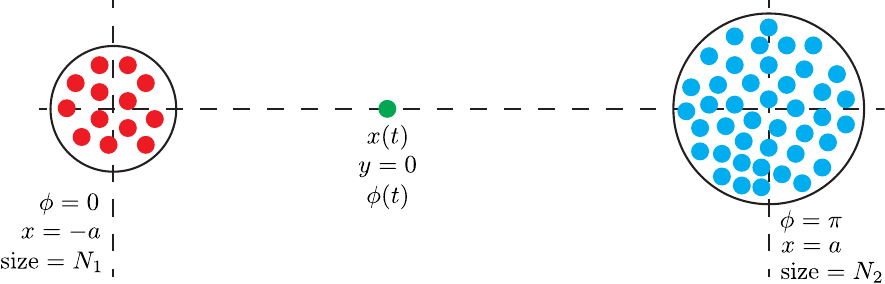}
\caption{
Schematic diagram of two large anti-phase clusters and a single vacillator. The two large clusters have centroids positioned at $(x,y) = (\pm a, 0)$, where $a =  \frac{R^*}{2} =  \frac{1}{2(1-J)}$ based on (\ref{eq:two_cluster_sep_distance}), and all swarmalators within each cluster have identical phases ($\phi=0$ or $\phi =\pi$). The large clusters contain $N_1$ and $N_2$ swarmalators, respectively. The single vacillator (green) has time-dependent position $(x(t),0)$ and time-dependent phase $\phi(t)$, with dynamics given by (\ref{eq:vacillator_x})-(\ref{eq:vacillator_phi}).
}
\label{fig:vacillator_schematic}
\end{figure}

Due to its analytical tractability, we consider here the dynamics of a single vacillator trapped between two anti-phase clusters. To yield a reduced model, we assume that the number of swarmalators in the clusters is sufficiently large so that the effect of the single vacillator on the two large clusters is negligible. We can therefore assume that the swarmalators within each large cluster have constant positions and constant phases. A schematic diagram summarizing the situation is shown in Fig.~\ref{fig:vacillator_schematic}. From the two-cluster reduction in Section~\ref{sec:anti-phase_clusters}, the asymptotic separation distance between the two clusters is approximated by (\ref{eq:two_cluster_sep_distance}). After an appropriate change of coordinates, we are able to specify the positions of the cluster centroids as $(\pm a, 0)$, where $a= \frac{R^*}{2} =  \frac{1}{2(1-J)}$. The vacillator has coordinates $(x,y,\phi)$. Since both clusters have centroid on the $x$-axis, the $y$-dynamics of the vacillator are of the form
\begin{equation}
\dot y \propto - y,
\end{equation}
meaning $y \to 0$ as $t\to \infty$, i.e., the vacillator converges toward the $x$-axis. We consider only the long-term dynamics, and, hence, set $y=0$ and only consider the $(x,\phi)$ dynamics. Without loss of generality, we assume all swarmalators in the cluster at $x=-a$ have phase $\phi = 0$, and all swarmalators in the cluster at $x=a$ have phase $\phi = \pi$. Let $N_1$ denote the number of swarmalators in the $x=-a$ cluster, and $N_2$ denote the number of swarmalators in the $x=a$ cluster. For sufficiently large $N$ we have $N_1 + N_2 \approx N$, and let $\alpha_1 = N_1/N$. By approximating the positions of swarmalators in the respective clusters by the cluster centroids we obtain reduced dynamical equations for the vacillator
\begin{align}
\dot x &= 1 - 2\alpha_1 - J \cos \phi - \frac{x + (1-2\alpha_1)a}{a^2 - x^2}, \label{eq:vacillator_x}   \\
\dot \phi & = \frac{\sin \phi}{a^2 - x^2} \left( K_1 \left(x + (1 - 2\alpha_1)a \right) - 2K_2 \cos\phi \left( (1-2 \alpha_1) x + a \right) \right). \label{eq:vacillator_phi}
\end{align}
Considering stationary solutions of the reduced system (\ref{eq:vacillator_x})-(\ref{eq:vacillator_phi}), the phase dynamics are stationary when $\sin \phi = 0$, i.e., $\phi = 0$ and $\phi =\pi$. For $\phi =0$, the $x$ dynamics are stationary when
\begin{equation} \label{eq:vacillator_absorption_1}
x = -a\left( \frac{2\sqrt{4 \alpha_1^2 a (a-1) +  \alpha_1 } - 1 }{ 4 \alpha_1 a -1 } \right) \approx -a,
\end{equation}
for $a \gg 0$ (equiv. $J\approx 1$). This stationary solution corresponds to absorption of the vacillator into the cluster at $x=-a$ with phase $\phi = 0$. Similarly, there is a stationary solution with $\phi = \pi$ and position
\begin{equation} \label{eq:vacillator_absorption_2}
x = a\left( \frac{2\sqrt{4 \alpha_2^2 a (a-1) +  \alpha_2 } - 1 }{ 4 \alpha_2 a -1 } \right) \approx a,
\end{equation}
where $\alpha_2 = N_2/N = 1- \alpha_1$, corresponding to absorption into the $x=a$ cluster with phase $\phi=\pi$. Both of these absorption stationary solutions are asymptotically stable for all relevant values of the parameters. There are also several other stationary solutions of the reduced system (\ref{eq:vacillator_x})-(\ref{eq:vacillator_phi}), corresponding to simultaneous solutions to
\begin{align}
\cos \phi &= \frac{2a}{2a - 1} \left( 1 - 2\alpha_1 - \frac{x + (1-2\alpha_1) a}{a^2 - x^2} \right),   \\
\cos \phi &= \frac{K_1}{2 K_2} \frac{x + (1-2 \alpha_1)a}{(1-2 \alpha_1)x + a},
\end{align}
which are equations for $x$-nullclines and $\phi$-nullclines, respectively. Simultaneous solutions to these nullcline equations satisfy a cubic equation in $x$
\begin{equation} \label{eq:cubic_x}
A_0 + A_1 x + A_2 x^2 + A_3 x^3 = 0
\end{equation}
where
\begin{align}
A_0 & = a^3 \beta_1 \left( \kappa (1- 2a) + 4 (a-1) \right),   \\
A_1 &= a^2\left( \kappa (1-2a) + 4 \left( (a-1)\beta_1^2 -1 \right) \right),   \\
A_2 &= -a \beta_1 \left( \kappa(1-2a)  + 4(a+1) \right),  \\ 
A_3 &= -\kappa (1-2a) - 4a \beta_1^2,
\end{align}
with $\beta_1 = 1- 2\alpha_1$ and $\kappa = K_1/K_2$. Therefore, there are either one, two, or three solutions to (\ref{eq:cubic_x}), each giving rise to a pair of stationary solutions to (\ref{eq:vacillator_x})-(\ref{eq:vacillator_phi}) due to the symmetry about $\phi=0$.

While the separation distance between the two clusters does not depend on the relative sizes of the clusters, we see that the dynamics of the vacillator does depend on the relative sizes of the two clusters. In Section~\ref{sec:vacillator_equal} we consider the case with equally sized clusters, i.e., $N_1 = N_2$, then discuss the effect of breaking this symmetry in Section~\ref{sec:vacillator_unequal}.

\subsection{Equally sized clusters} \label{sec:vacillator_equal}

\begin{figure}[tbp]
\centering
\includegraphics[width=\columnwidth]{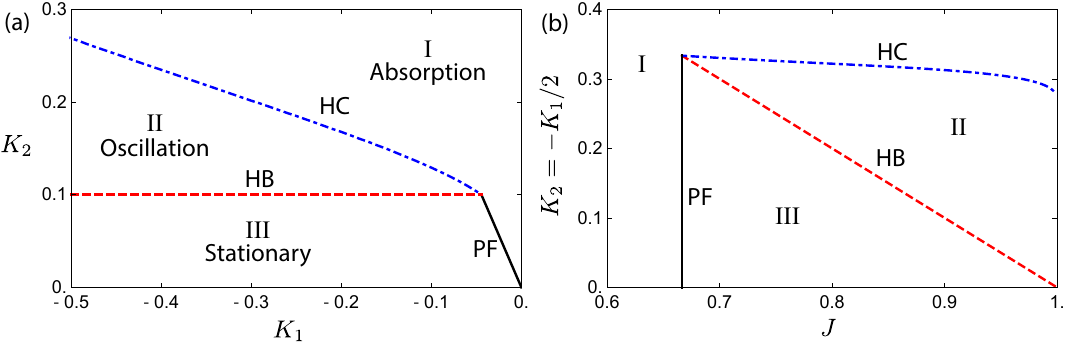}
\caption{
Bifurcation diagrams for the reduced vacillator system (\ref{eq:vacillator_x})-(\ref{eq:vacillator_phi}. (a)~Bifurcations in the $K_1$-$K_2$ plane for fixed $J=0.9$.  Subcritical pitchfork (PF, solid black), supercritical Hopf (HB, dashed red) and a pair of simultaneous heteroclinic (HC dot-dashed blue) bifurcations separate the regions I, II and III. (b)~Bifurcations with varying $J$ and $K_2$, keeping $K_1 = -2K_2$.
}
\label{fig:two_cluster_one_vacillator_bif_diagrams}
\end{figure}

\begin{figure}[tbp]
\centering
\includegraphics[width=\columnwidth]{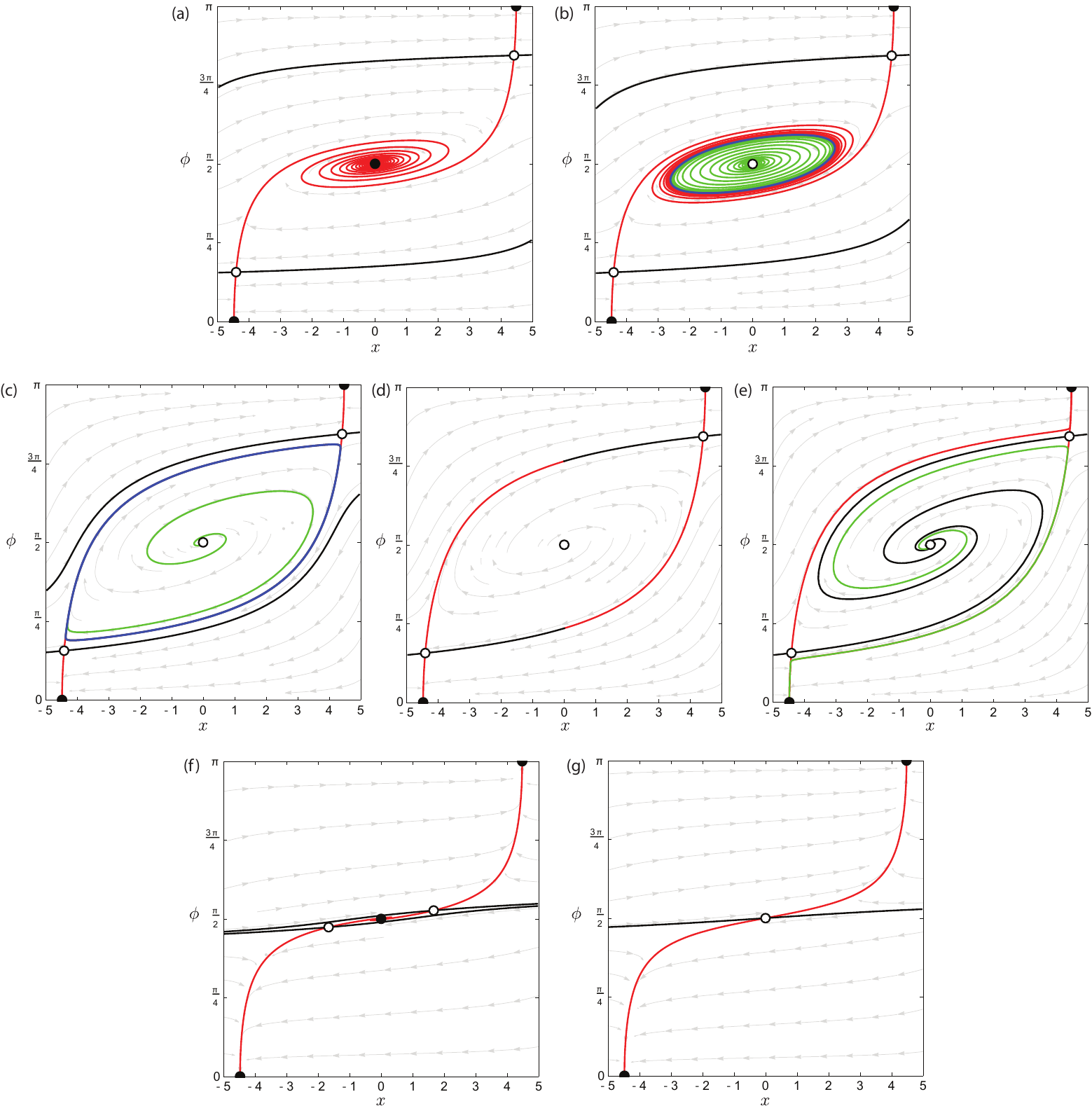}
\caption{
Phase portraits for the reduced vacillator dynamics (\ref{eq:vacillator_x})-(\ref{eq:vacillator_phi}) for various values of $K_1$ and $K_2$, with $J=0.9$ and $\alpha_1 = 0.5$ kept fixed for all plots. Stable (filled circles) and unstable (open circles) stationary points are shown together with stable (black) and unstable (red) manifolds associated with saddle equilibria. Streamlines are shown in gray with arrows. Periodic orbits are shown in blue, and the trajectory of an initial condition close to the equilibrium $(0,\pi/2)$ is shown in green. 
(a)~$K_1 = -0.16$, $K_2 = 0.08$ (region III), (b)~$K_1 = -0.24$, $K_2 = 0.12$ (region II), (c)~$K_1 = -0.56$, $K_2 = 0.28$ (region II), (d)~$K_1 = -0.6252$, $K_2 = 0.3126$ (pair of heteroclinic connections HC), (e)~$K_1 = -0.68$, $K_2 = 0.34$ (region I), (f)~$K_1 = -0.04$, $K_2 = 0.08$ (region III), (g)~$K_1 = -0.028$, $K_2 = 0.08$ (region I).
}
\label{fig:two_cluster_one_vacillator_pplane}
\end{figure}

In the case of equally sized clusters, $N_1 = N_2$, the equations simplify significantly. In this case $\alpha_1 = 1/2$ and $\beta_1 = 0$. The cubic equation (\ref{eq:cubic_x}) becomes
\begin{equation}
x \left( a^2 \left(  \kappa (1-2a) - 4 \right) - \kappa(1-2a) x^2 \right) = 0
\end{equation}
with roots at $x=0$ and 
\begin{equation} \label{eq:two_cluster_one_vacillator_saddle_x}
x = \pm \frac{a\sqrt{4+(2a-1) \kappa}}{\sqrt{ (2a-1) \kappa}}.
\end{equation}
The corresponding stationary solutions have $\phi$ satisfying
\begin{equation}  \label{eq:two_cluster_one_vacillator_phi}
\cos \phi = \frac{\kappa x}{2 a}.
\end{equation}
Stability analysis shows that the stationary solutions corresponding to (\ref{eq:two_cluster_one_vacillator_saddle_x}) are saddles for all relevant parameter ranges.
For the stationary solution with $x=0$, (\ref{eq:two_cluster_one_vacillator_phi}) yields the symmetric pair $\phi = \pm \pi/2$. At $(x,\phi) = (0, \pm \pi/2)$ the Jacobian of the reduced system (\ref{eq:vacillator_x})-(\ref{eq:vacillator_phi}) is equal to
\begin{equation}
\mathcal{J} = \frac{1}{a^2} \begin{pmatrix}
-1 & \pm a(a-1/2) \\
\pm K_1 & 2aK_2
\end{pmatrix}.
\end{equation}
Letting $\tau$ and $\Delta$ denote the trace and determinant of $\mathcal{J}$, respectively, we obtain
\begin{align}
\tau &= \frac{1}{a^2} (2 a K_2 -1),   \\
\Delta &= \frac{1}{2 a^3} \left( (1-2a)K_1 -4 K_2 \right).
\end{align}
Therefore, the stationary solutions $(x,\phi) = (0, \pm \pi/2)$ are stable provided $\tau < 0$ and $\Delta >0$, i.e.,
\begin{align}
K_2 &< \frac{1}{2a} = 1 - J, \quad \text{and} \\
K_2 &< \frac{1-2a}{4} K_1 = \frac{J}{4(J-1)} K_1.
\end{align}
This stable region is shown in the $K_1$-$K_2$ plane for $J=0.9$ in Fig.~\ref{fig:two_cluster_one_vacillator_bif_diagrams}(a) by the region III. Similarly, this stable region is shown as region III in the $J$-$K_2$ plane in Fig.~\ref{fig:two_cluster_one_vacillator_bif_diagrams}(b), where we restrict to the line $K_1 = -2K_2$. In this region of the parameter space the vacillator is in stable equilibrium at the midpoint between the two clusters, and has phase $\pm \pi/2$, i.e., out of phase by $\pi/2$ from the two anti-phase clusters. A typical phase plane in this region is shown in Fig.~\ref{fig:two_cluster_one_vacillator_pplane}(a) for $K_1 = -0.16$, $K_2 = 0.08$ and $J=0.9$. These parameters correspond to cluster positions $x = \pm a$ with $a=5$. The domain of interest is $(x,\theta) \in [-5,5]\times [0,2\pi)$. Only the range $\theta \in [0,\pi]$ is shown, since the range $\theta \in [\pi,2\pi]$ is essentially the same, except reflected about $x=0$ ($x \mapsto -x$). There are three stable equilibria (closed circles), two corresponding to the absorption states (\ref{eq:vacillator_absorption_1}) and (\ref{eq:vacillator_absorption_2}), and the vacillator state $(x,\phi) = (0, \pi/2)$. The basins of attraction for these stable equilibria are separated by the stable manifolds (solid black curves) associated with the symmetric saddle equilibria (open circles) given by (\ref{eq:two_cluster_one_vacillator_saddle_x}) and (\ref{eq:two_cluster_one_vacillator_phi}).

 Along the plane corresponding to $\tau = 0$, i.e., $K_2 = 1-J$ (dashed red lines in Fig.~\ref{fig:two_cluster_one_vacillator_bif_diagrams}), a supercritical Hopf bifurcation (HB) occurs, such that the equilibrium point at $(x, \theta) = (0,\pi/2)$ loses stability, and a stable limit cycle emerges. This limit cycle corresponds to persistent wavering of the vacillator between the two large clusters (wavering both in space and in phase), and characterizes region II in Fig.~\ref{fig:two_cluster_one_vacillator_bif_diagrams}. This Hopf bifurcation is demonstrated by the transition between the phase planes Fig.~\ref{fig:two_cluster_one_vacillator_pplane}(a) and Fig.~\ref{fig:two_cluster_one_vacillator_pplane}(b). The limit cycle is shown as the blue curve, and is approached on the inside by the green solution curve and on the outside by the unstable manifolds (red) associated with the pair of saddle equilibria. The stable manifolds form a separatrix dividing the domain into initial conditions that are attracted to the limit cycle and initial conditions that are absorbed into one of the large clusters.
 
Moving away from the Hopf bifurcation, the limit cycle amplitude increases (cf. Fig.~\ref{fig:two_cluster_one_vacillator_pplane}(b) and Fig.~\ref{fig:two_cluster_one_vacillator_pplane}(c)). Along a critical surface in $K_1$-$K_2$-$J$ parameter space, the unstable and stable manifolds of the symmetric saddle points merge in a pair of heteroclinic connections (HC) (dot-dashed blue curves in Fig.~\ref{fig:two_cluster_one_vacillator_bif_diagrams}). A phase portrait at a critical value is shown in Fig.~\ref{fig:two_cluster_one_vacillator_pplane}(d), where the unstable (red) and stable (black) manifolds coincide.  Beyond the heteroclinic connection there is no limit cycle solution (cf. Fig.~\ref{fig:two_cluster_one_vacillator_pplane}(e)), and the only stable solutions are the absorption states (corresponding to region I in Fig.~\ref{fig:two_cluster_one_vacillator_bif_diagrams}).

 Along the surface corresponding to $\Delta = 0$, i.e., $K_2 =  \frac{J}{4(J-1)} K_1$ (solid black lines in Fig.~\ref{fig:two_cluster_one_vacillator_bif_diagrams}), a subcritical pitchfork bifurcation (PF) occurs, such that the two symmetric saddles given by (\ref{eq:two_cluster_one_vacillator_saddle_x}) and (\ref{eq:two_cluster_one_vacillator_phi}) coalesce with the stable stationary solution at $x=0$, resulting in a single saddle equilibrium at $x=0$ beyond the bifurcation. This is demonstrated by the transition from Fig.~\ref{fig:two_cluster_one_vacillator_pplane}(f) to Fig.~\ref{fig:two_cluster_one_vacillator_pplane}(g). Thus, beyond the bifurcation the vacillator is absorbed into one of the two clusters (region I in Fig.~\ref{fig:two_cluster_one_vacillator_bif_diagrams}).

\subsubsection{Reduced model compared to full model} 
 
 When simulating the full model with one vacillator, we begin with an equilibrium solution of the full model with two large clusters of a given size, the clusters have mean phases $\Phi_1=0$ and $\Phi_2 = \pi$, and are centered such that the mean position of all swarmalators is at the origin. A swarmalator is then added between the two clusters, with a random position $x$ close to zero and a random phase $\phi$ close to $\pi/2$. After initial seeding, the full model is run for a transient time of 1,000 time units before data is recorded.
 
  To detect bifurcations in the full model (\ref{eq:swarmalator_space})-(\ref{eq:swarmalator_phase}), and to compare with the reduced model (\ref{eq:vacillator_x})-(\ref{eq:vacillator_phi}), we compute the minimum difference between the phase of the vacillator and the time-averaged phase of each cluster, i.e.,
 \begin{equation} \label{eq:vacillator_phase_diff}
 \min_{t>0} \min_{j=1,2} |\phi(t) - \bar\Phi_j|,
\end{equation}  
where $\phi(t)$ is the phase of the vacillator, $\bar \Phi_j = \arg\left( \frac{1}{T} \int_0^T \exp(i \Phi_j(t))dt \right)$ is the mean phase of cluster $j$, and the difference accounts for arithmetic modulo $2\pi$. In cases where the vacillator is stationary, e.g., Fig.~\ref{fig:two_cluster_one_vacillator_pplane}(a,f), the minimum phase difference (\ref{eq:vacillator_phase_diff}) is close to $\pi/2$ (it is exacltly $\pi/2$ in the reduced model (\ref{eq:vacillator_x})-(\ref{eq:vacillator_phi})). For a limit cycle solution, e.g., Fig.~\ref{fig:two_cluster_one_vacillator_pplane}(b,c), the minimum phase difference (\ref{eq:vacillator_phase_diff}) is between $0$ and $\pi/2$. When the vacillator is absorbed into one of the two clusters, e.g. Fig.~\ref{fig:two_cluster_one_vacillator_pplane}(e,g), the minimum phase difference (\ref{eq:vacillator_phase_diff}) is zero, because it has identical phase with the cluster that it has been absorbed into. Thus, the minimum phase difference (\ref{eq:vacillator_phase_diff}) can detect the bifurcations observed in Fig.~\ref{fig:two_cluster_one_vacillator_bif_diagrams} and Fig.~\ref{fig:two_cluster_one_vacillator_pplane}, as well as measuring the amplitude of any limit cycle solutions (larger amplitude limit cycles yield small values of the minimum phase difference (\ref{eq:vacillator_phase_diff})).

\begin{figure}[tbp]
 \centering
 \includegraphics[width=\columnwidth]{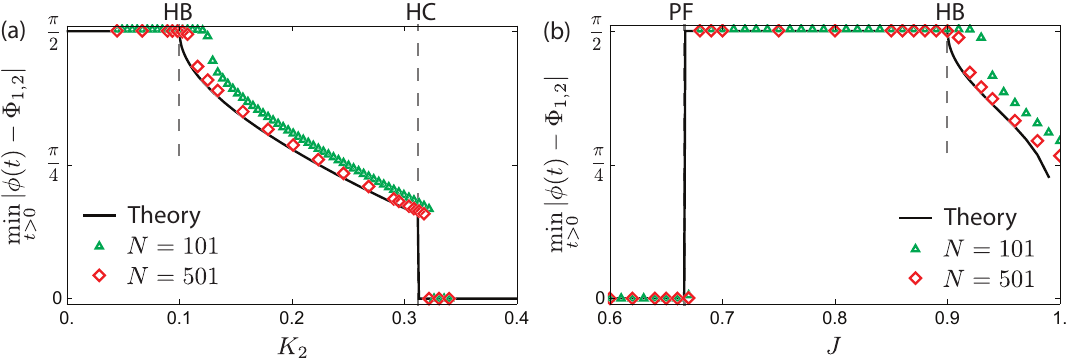	}
 \caption{
The minimum phase difference between the vacillator and the two clusters (\ref{eq:vacillator_phase_diff}) shown for the reduced model (\ref{eq:vacillator_x})-(\ref{eq:vacillator_phi}) (solid black) and full model (\ref{eq:swarmalator_space})-(\ref{eq:swarmalator_phase}) ($N=101$: green triangles and $N=501$: red diamonds) demonstrates bifurcations (HB, HC, PF) in the dynamics. (a)~Varying $K_2$ with $J=0.9$ and $K_1 = -2K_2$. (b)~Varying $J$ with $K_1=-0.2$ and $K_2 = 0.1$. 
 }
 \label{fig:two_cluster_w_vacillator_full_vs_reduced}
 \end{figure}

 Fig.~\ref{fig:two_cluster_one_vacillator_bif_diagrams}(b) shows that for fixed $J>2/3$, and maintaining $K_1 = -2K_2$, the reduced model (\ref{eq:vacillator_x})-(\ref{eq:vacillator_phi}) predicts two bifurcations to occur as $K_2$ is varied. Increasing $K_2$ from zero, first there is a Hopf bifurcation dividing region III (stationary equilibrium) and region II (limit cycle solution), then there is a heteroclinic bifurcation dividing region II and region I (absorption). Fig.~\ref{fig:two_cluster_w_vacillator_full_vs_reduced}(a) shows that the reduced model accurately captures the dynamics and bifurcations that occur in the full system with $J=0.9$ and $K_2$ varied (with $K_1 = -2K_2$). The reduced model (\ref{eq:vacillator_x})-(\ref{eq:vacillator_phi}), shown as the solid black curve, predicts the Hopf bifurcation at $K_2 = 0.1$, such that the minimum phase difference (\ref{eq:vacillator_phase_diff}) decreases from $\pi/2$ when a limit cycle emerges. This shift is closely matched by the numerical simulations of the full model  (\ref{eq:swarmalator_space})-(\ref{eq:swarmalator_phase}), where green triangles show results for $N=101$ swarmalators ($50$ in each cluster) and red diamonds show results for $N=501$ swarmalators ($250$ in each cluster). The reduced model better describes the case with $N=501$, which is expected because the reduced model assumes infinitely many swarmalators in each of the clusters. For the heteroclinic bifurcation, this occurs at $K_2 = 0.3126$ in the reduced model, which agrees well with the bifurcation observed in the full system, such that the vacillator is absorbed into one of the clusters and the minimum phase difference (\ref{eq:vacillator_phase_diff}) becomes zero. Again, the reduced model is more accurate for the case with $N=501$ compared to $N=101$, as expected.
 
 To show that the subcritical pitchfork bifurcation observed in the reduced model agrees with the full model, we keep $K_1 = -0.2$ and $K_2 =0.1$ fixed, and vary $J$. This corresponds to traversing the horizonal line through $K_2 = 0.1$ in Fig.~\ref{fig:two_cluster_one_vacillator_bif_diagrams}(b). The reduced model (\ref{eq:vacillator_x})-(\ref{eq:vacillator_phi}) predicts that as $J$ is increased, a subcritical pitchfork bifurcation occurs at $J= 2/3$, giving rise to a stable stationary solution. Fig.~\ref{fig:two_cluster_w_vacillator_full_vs_reduced}(b) shows that this is accurate for the dynamics of the full model. The jump in the minimum phase difference (\ref{eq:vacillator_phase_diff}) from $0$ to $\pi/2$ close to $J=2/3$ indicated the birth of a stable stationary solution. As $J$ is further increased, the Hopf bifurcation occurs at $J=0.9$, such that a limit cycle solution emerges and the minimum phase difference begins to decrease, which is again reflected in the dynamics of the full model. As expected, the reduced model is more accurate for the case with $N=501$ compared to $N=101$.

\subsection{Unequal cluster sizes} \label{sec:vacillator_unequal}
Breaking the symmetry $N_1 = N_2$ breaks many of the dynamical symmetries that we observe, such as the symmetries in the absorption states (\ref{eq:vacillator_absorption_1})-(\ref{eq:vacillator_absorption_2}) as well as the symmetry in the cubic equation (\ref{eq:cubic_x}) which defines the non-absorption equilibria. Breaking the size symmetry also affects the structurally unstable bifurcations that we observe (simultaneous heteroclinic connections and pitchfork bifurcations).

\begin{figure}[tbp]
\centering
\includegraphics[width=\columnwidth]{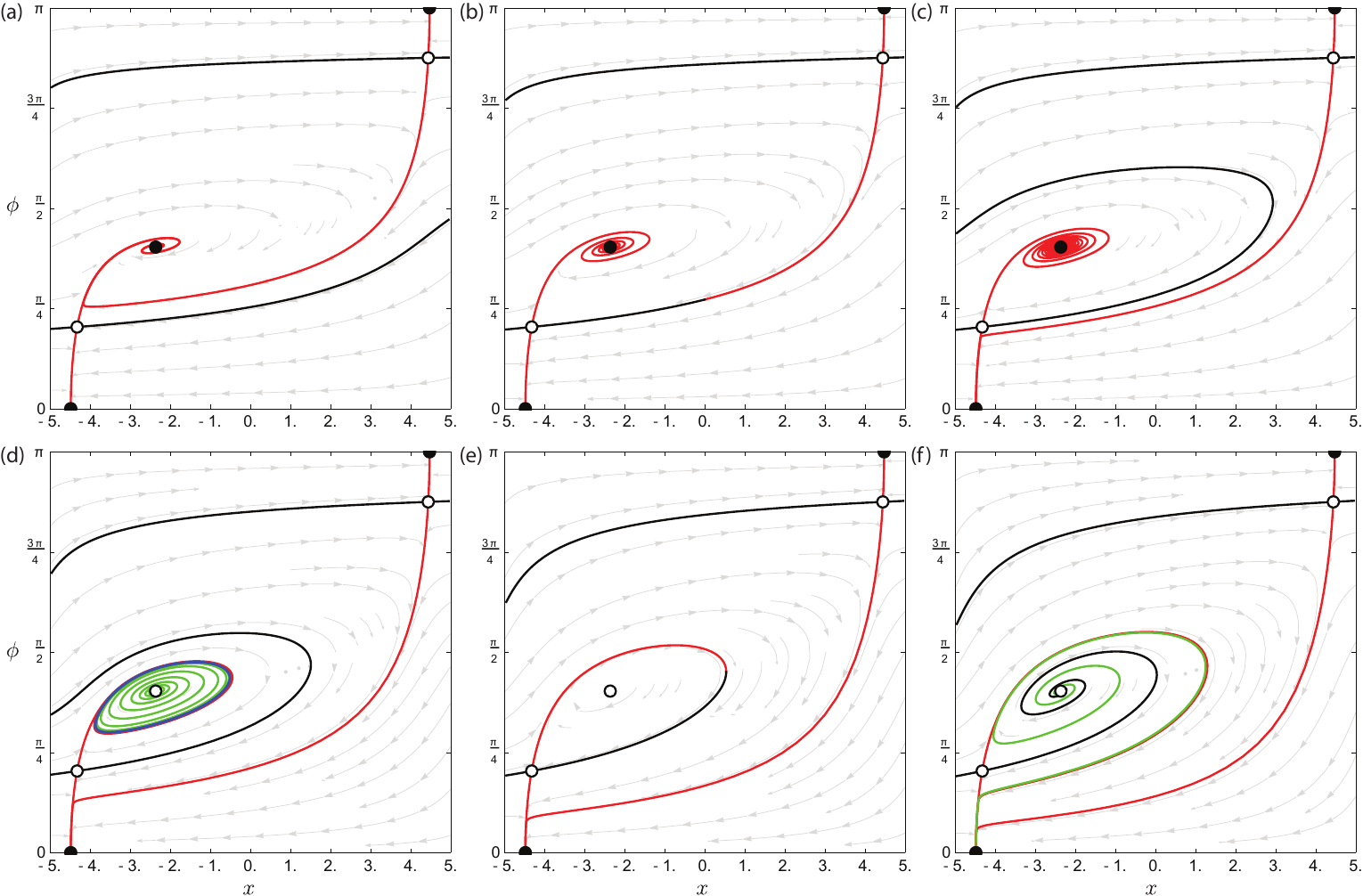}
\caption{
Phase portraits for the reduced vacillator dynamics (\ref{eq:vacillator_x})-(\ref{eq:vacillator_phi}) for a range of $K_2$ values with $J=0.9$, $K_1 = -2K_2$ and $\alpha_1 = 0.4$ kept fixed. Stable (filled circles) and unstable (open circles) equilibria are shown together with stable (black) and unstable (red) manifolds associated with saddle equilibria. Streamlines are shown in gray with arrows. Periodic orbits are shown in blue, and the trajectory of an initial condition close to the equilibrium at $(x,\phi)=(-2.3577,1.2663)$ is shown in green. (a)~$K_2=0.1$, (b)~$K_2=0.1331$ (heteroclinic connection), (c)~$K_2 = 0.15$, (d)~$K_2=0.2$, (e)~$K_2 = 0.2581$ (homoclinic connection), and (f)~$K_2=0.3$.
}
\label{fig:unequal_vacillator_pplanes}
\end{figure}

\begin{figure}[tbp]
\centering
\includegraphics[width=0.5\columnwidth]{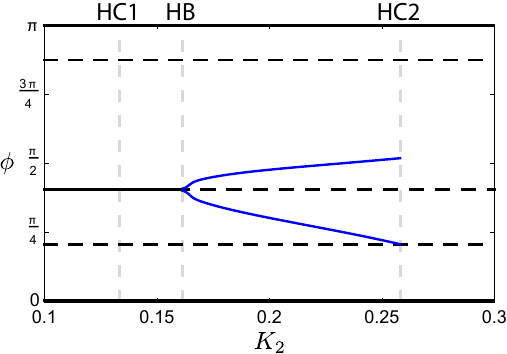}
\caption{
Bifurcation diagram for the reduced vacillator dynamics (\ref{eq:vacillator_x})-(\ref{eq:vacillator_phi}) with $K_2$ varying and $J=0.9$, $K_1 = -2K_2$ and $\alpha_1 = 0.4$ kept fixed. Stable (solid) and unstable (dashed) equilibria are shown together with maximum and minimum values of periodic orbits (blue). At $K_2 = 1331$ a heteroclinic bifurcation occurs (HC1, cf. Fig.~\ref{fig:unequal_vacillator_pplanes}(b)). At $K_2 = 0.1613$ a supercritical Hopf bifurcation occurs (HB). At $K_2 = 0.2581$ a homoclinic bifurcation occurs (HC2, cf. Fig.~\ref{fig:unequal_vacillator_pplanes}(e)).
}
\label{fig:unequal_vacillator_bif}
\end{figure}

The bifurcations that occur with $J=0.9$, $K_1 = -2K_2$ and $\alpha_1 = 0.4$ kept fixed, with $K_2$ varying are shown in Fig.~\ref{fig:unequal_vacillator_pplanes} using phase portraits, and are summarized in the bifurcation diagram Fig.~\ref{fig:unequal_vacillator_bif}. At $K_2=0.1331$ a heteroclinic bifurcation occurs, such that the stable manifold from the left saddle equilibrium and the unstable manifold from the right saddle equilibrium coincide. This results in a sudden reduction in the basin of attraction for the stable vacillator equilibrium at $(x,\phi)=(-2.3577,1.2663)$ (compare Fig.~\ref{fig:unequal_vacillator_pplanes}(a) with Fig.~\ref{fig:unequal_vacillator_pplanes}(c)), meaning more random initial conditions will be absorbed into one of the clusters. At $K_2 = 0.1613$ a supercritical Hopf bifurcation occurs, giving rise to a stable limit cycle (shown in blue in Fig.~\ref{fig:unequal_vacillator_pplanes}(d)). As $K_2$ increases, the amplitude of the limit cycle grows, and at $K_2 = 0.2581$ the limit cycle is destroyed via a homoclinic bifurcation (cf. Fig.~\ref{fig:unequal_vacillator_pplanes}(e)). For $K_2>0.2581$, all initial conditions result in absorption into one of the clusters, with most initial conditions being absorbed into the smaller cluster at $x\approx -5$ with $\phi =0$.

The subcritical pitchfork bifurcation observed for equal sized clusters (cf. Fig.~\ref{fig:two_cluster_one_vacillator_bif_diagrams} and Fig.~\ref{fig:two_cluster_one_vacillator_pplane}(f,g)) is also structurally unstable and upon perturbation becomes a saddle node bifurcation, such that the stable equilibrium and one of the saddle equilibria coalesce and annihilate at bifurcation.

\subsection{Multiple vacillators}

We note that a reduction similar to (\ref{eq:vacillator_x})-(\ref{eq:vacillator_phi}) can be performed in the case of multiple vacillators. For example, for the case with four vacillators shown in Fig.~\ref{fig:four_vacillator}, the large clusters can be considered stationary with constant phases, leaving dynamics for the four vacillators, i.e., a 12-dimensional system. However, such a reduction is challenging. As in (\ref{eq:vacillator_x})-(\ref{eq:vacillator_phi}), the vacillator-cluster interactions are $\mathcal{O}(1)$, but the vacillator-vacillator interactions will be $\mathcal{O}(1/N)$. It is necessary to assume that $N$ is large so that the effect of the vacillators on the clusters can be neglected, but large $N$ results in a stiff system of ODE's that is challenging to solve numerically with high precision.

\section{Higher harmonics in the coupling function} \label{sec:HH}

As expected, including higher harmonics in the coupling function yields multiple phase clusters. For simplicity, in this section we consider only a single (higher) harmonic in the phase dynamics coupling function, rather than combinations of higher harmonics. As such, we consider phase dynamics given by
\begin{equation}
\dot{\phi}_i = \frac{K}{N} \sum_{j=1, \, j\neq i}^N \frac{\sin\left( m (\phi_j - \phi_i ) \right)}{|\bm{x}_j - \bm{x}_i|} , \label{eq:swarmalator_phase_HH}
\end{equation}
where $m$ is the chosen harmonic. This phase dynamics is combined with the same spatial dynamics (\ref{eq:swarmalator_space}) as used previously. Choosing $K=K_2$ and $m=2$ recovers the dynamics (\ref{eq:swarmalator_phase}) in the case that $K_1=0$. 

Considering equilibria of (\ref{eq:swarmalator_phase_HH}), we see that $\dot{\phi}_i = 0$ if $\sin\left( m (\phi_j - \phi_i ) \right) = 0$ for all $i$ and $j$. This is satisfied if the phases are of the form
\begin{equation} \label{eq:HH_eq}
\phi_i = k_i\frac{2\pi}{m} + \Theta,
\end{equation}
where $k_i\in \lbrace 0,...,m-1 \rbrace$ and $\Theta$ is a common offset. It is therefore typical that the dynamics (\ref{eq:swarmalator_space})-(\ref{eq:swarmalator_phase_HH}) yields $m$ distinct equally distributed phases. Considering the stability of these phase equilibria, the Jacobian of the phase dynamics (\ref{eq:swarmalator_phase_HH}) is given by
\begin{equation}
\left(\mathcal{J}(\bm{\phi})\right)_{ij} = \frac{\partial \dot\phi_i}{\partial \phi_j}  = 
\frac{mK}{N} 
\begin{cases} 
- \sum_{k \neq i} \frac{ \cos(m(\phi_k - \phi_i))}{|\bm{x}_k - \bm{x}_i|}, & i=j, \\
\frac{ \cos(m(\phi_j - \phi_i))}{|\bm{x}_j - \bm{x}_i|}, & i \neq j.
\end{cases}
\end{equation}
At an equilibrium state of the form (\ref{eq:HH_eq}), this Jacobian is equal to
\begin{align}
\left(\mathcal{J}(\bm{\phi}^*)\right)_{ij}  &= 
-\frac{mK}{N} 
\begin{cases} 
 \sum_{k \neq i} \frac{1 }{|\bm{x}_k - \bm{x}_i|}, & i=j,  \\
-\frac{1 }{|\bm{x}_j - \bm{x}_i|}, & i \neq j,
\end{cases} \nonumber \\
&= - \frac{mK}{N}  \mathcal{L}_{ij},  \label{eq:HH_Jacobian}
\end{align}
where $\mathcal{L}$ is the graph Laplacian of the weighted undirected network with adjacency matrix $A_{ij} = \frac{1 }{|\bm{x}_j - \bm{x}_i|}$. As such, if $K>0$ then the phase equilibria (\ref{eq:HH_eq}) are stable under the phase dynamics (\ref{eq:swarmalator_phase_HH}) for any fixed spatial configuration, and, as discussed in Section~\ref{sec:static_single_cluster}, the spatial dynamics (\ref{eq:swarmalator_space}) are guaranteed to reach a stable equilibrium corresponding to a local minimum of the interaction potential $U(\bm{x})$ defined via (\ref{eq:space_potential}).

\begin{figure}[tbp]
\centering
\includegraphics[width=\columnwidth]{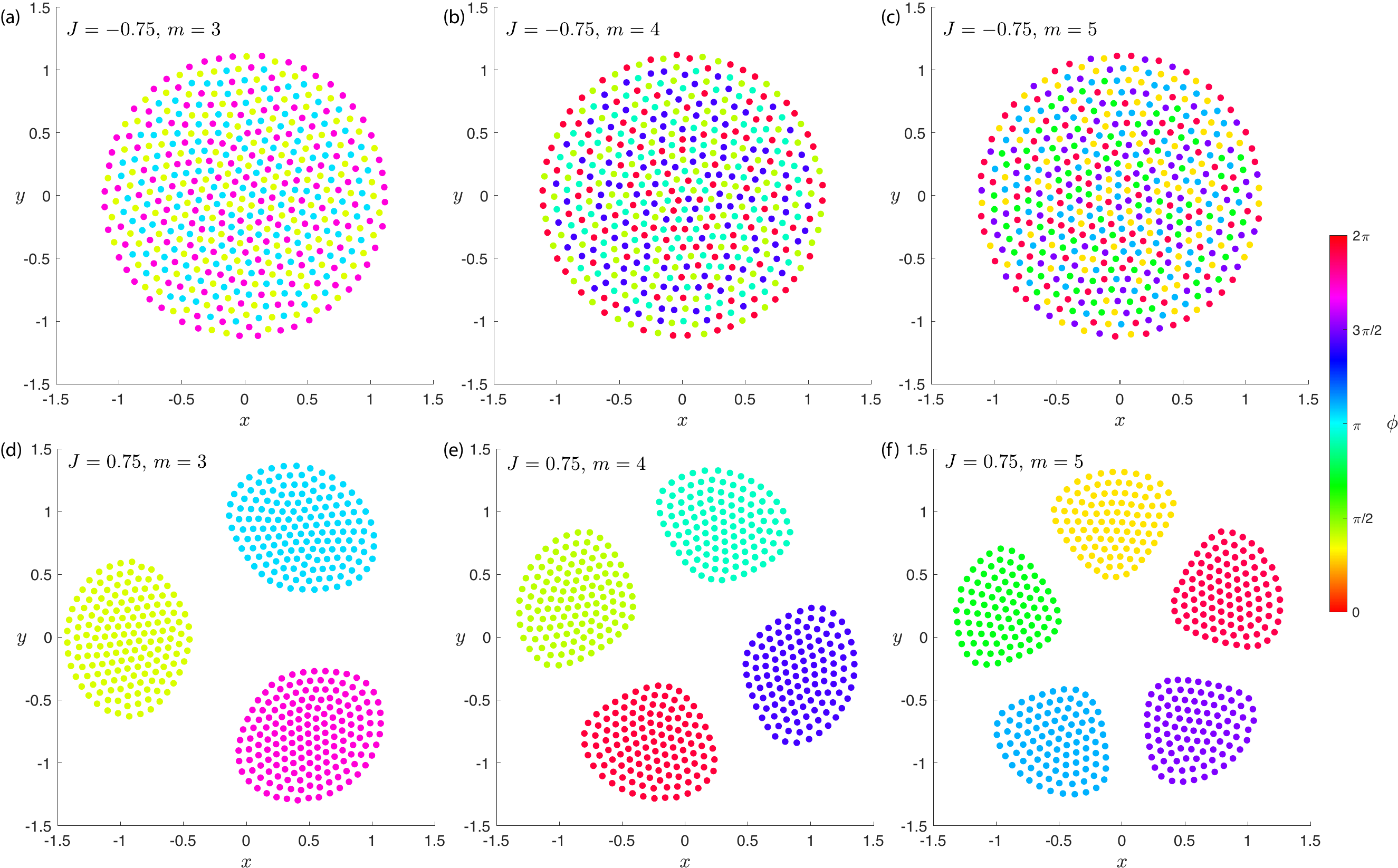}
\caption{
Static states for the higher harmonic swarmalator dynamics (\ref{eq:swarmalator_space})-(\ref{eq:swarmalator_phase_HH}) with $K=1$ and $N=500$. Top row: $J=-0.75$ and (a)~$m=1$, (b)~$m=2$, (c)~$m=3$. Bottom row: $J=0.75$ and (d)~$m=1$, (e)~$m=2$, (f)~$m=3$.
}
\label{fig:higher_harmonics}
\end{figure}

In the case $J<0$, the swarmalators form a single spatial cluster with $m$ distinct phases, akin to those in Fig.~\ref{fig:2_phase_1_cluster}. This is demonstrated in the top row of Fig.~\ref{fig:higher_harmonics} for $J=-0.75$ and $m=3,4,5$.

In the case $J>0$, the swarmalators arrange themselves into $m$ spatial clusters, with each cluster having a unique phase, similar to the two-cluster cases shown in Fig.~\ref{fig:2_phase_2_cluster}. This is demonstrated in the bottom row of Fig.~\ref{fig:higher_harmonics} for $J=0.75$ and $m=3,4,5$. Future work should focus on the spatial arrangements of these clustered states.

We remark that the state with two anti-phase clusters that occurs for $m=2$ (Fig.~\ref{fig:2_phase_2_cluster}) also arises in the swarmalator model with attractive local coupling and repulsive distant coupling \cite{SarEtAl2022}. However, the states with multiple clusters that arise from $m\geq 3$ in (\ref{eq:swarmalator_phase_HH}), i.e., those in Fig.~\ref{fig:higher_harmonics}, require higher harmonic coupling and do not occur in the model \cite{SarEtAl2022}.

\section{Conclusions} \label{sec:conclusions}

As a step toward studying general phase coupling functions in systems of swarmalators we have considered the inclusion of higher harmonic phase coupling. We have found novel clustered states that do not occur without higher harmonic coupling, including states that are clustered both spatially and by phase, and states that are clustered only by phase. We have determined their parametric stability regions by reducing the stability problem to that of purely phase dynamics.

In the case of two anti-phase spatial clusters, we have used mean field reduction to determine the spatial separation of the clusters, and have verified our theoretical result when compared to the full model.

We have also studied novel states with two large anti-phase clusters and a small number of vacillators that waver between them. By considering a mean-field reduction we are able to reduce the dynamics of the system with one vacillator to a two-dimensional differential equation, which allows for a detailed exploration of its bifurcation structure. We show that the vacillator transitions between stationary to oscillatory dynamics via a Hopf bifurcation, and is absorbed into one of the two clusters upon a heteroclinic bifurcation. We have shown that the dynamics of the reduced model agrees excellently with the full swarmalator model.

Future work should focus on unraveling the complex stability and bifurcation properties of the swarmalator model with combinations of higher harmonics, forming higher accuracy Fourier series truncations of general coupling functions. Here we have considered combined first and second harmonics, and then individual higher harmonics, but the dynamics will become more complex if several higher harmonics are considered simultaneously, including the possibility of even more complex vacillator dynamics.

We have focused on clustered, mostly stationary, states. We have also found many complex non-stationary attracting states, and transitions between them. For instance, for fixed $K_1<0$ and $J>0$, we have shown that the two cluster state is stable for $K_2>-K_1/2$, but begins to fragment for $K_2<-K_1/2$, eventually forming either a phase wave or splintered phase wave at $K_2=0$ \cite{OKeeffeEtAl2017}. More work is needed to understand these complex transitions.

\section*{Acknowledgments}
I would like to thank Ankith Das and Nikolas Petranovic for our insightful discussions and efforts in their respective summer research programmes.

\bibliographystyle{siamplain}
\bibliography{swarmalators-1}

\end{document}